\documentclass[useAMS,usenatbib]{mnras}

\usepackage{graphicx}
\usepackage{amssymb,multirow}
\usepackage[fleqn]{amsmath}
\usepackage{mathrsfs}
\usepackage{epstopdf,enumerate,url,color}
\usepackage{bm}
\usepackage{subcaption}
\usepackage{enumitem}
\usepackage{enumerate}
\usepackage{tabularx}

\title[Secular Chaos]{Formation of Hot Jupiters through Secular Chaos and Dynamical Tides}
\author[Teyssandier et al.]{Jean Teyssandier\thanks{E-mail: jt553@cornell.edu}, Dong Lai \& Michelle Vick\\ 
Department of Astronomy, Cornell Center for Astrophysics and Planetary Science, Cornell University, Ithaca, NY 14853, USA
}

\newcommand{\dd}[2]{\frac{\mathrm{d} #1}{\mathrm{d} #2}}

\newcommand{\Op}{\Omega_{\rm p}}
\newcommand{\Os}{\Omega_{\rm *}}
\newcommand{\Mp}{M_{\rm p}}
\newcommand{\Rp}{R_{\rm p}}
\newcommand{\Msun}{M_{\odot}}
\newcommand{\Ms}{M_*}
\newcommand{\Rsun}{R_{\odot}}
\newcommand{\Rs}{R_{\rm *}}
\newcommand{\Mj}{M_{\rm J}}
\newcommand{\Rj}{R_{\rm J}}

\newcommand{\Cp}{C_{\rm P}}
\newcommand{\Sp}{S_{\rm p}}
\newcommand{\Spv}{\bm{S}_{\rm p}}
\newcommand{\Sphat}{\bm{\hat{S}}_{\rm p}}
\newcommand{\Ip}{I_{\rm p}}
\newcommand{\kp}{k_{\rm p}}

\newcommand{\Ssv}{\bm{S}_{\rm *}}
\newcommand{\Sshat}{\bm{\hat{S}}_{\rm *}}
\newcommand{\Is}{I_{\rm *}}
\newcommand{\ks}{k_{\rm *}}
\newcommand{\costsl}{\cos\theta_{\rm sl}}
\newcommand{\rt}{r{}_{\rm tide}}
\newcommand{\dtl}{\Delta t_{\rm L}}
\newcommand{\ktwop}{k_{\rm 2p}}
\newcommand{\Lhat}{\bm{\hat{L}}}

\begin{document}
\maketitle

\begin{abstract}
The population of giant planets on short-period orbits can potentially be explained by some flavours of high-eccentricity migration. In this paper we investigate one such mechanism involving ``secular chaos'', in which secular interactions between at least three giant planets push the inner planet to a highly eccentric orbit, followed by tidal circularization and orbital decay. In addition to the equilibrium tidal friction, we incorporate dissipation due to dynamical tides that are excited inside the giant planet. Using the method of Gaussian rings to account for planet-planet interactions, we explore the conditions for extreme eccentricity excitation via secular chaos and the properties of hot Jupiters formed in this migration channel. Our calculations show that once the inner planet reaches a sufficiently large eccentricity, dynamical tides quickly dissipate the orbital energy, producing an eccentric warm Jupiter, which then decays in semi-major axis through equilibrium tides to become a hot Jupiter. Dynamical tides help the planet avoid tidal disruption, increasing the chance of forming a hot Jupiter, although not all planets survive the process. We find that the final orbital periods generally lie in the range of 2-3 days, somewhat shorter than those of the observed hot Jupiter population. We couple the planet migration to the stellar spin evolution to predict the final spin-orbit misalignments. The distribution of the misalignment angles we obtain shows a lack of retrograde orbits compared to observations. Our results suggest that high-eccentricity migration via secular chaos can only account for a fraction of the observed hot Jupiter population.
\end{abstract}


\section{Introduction}
\label{sec:intro}

About $1\%$ of Sun-like stars contain giant planets with periods
shorter than 10 days, the so-called hot Jupiters \citep[see e.g,][and references therein]{wf15,dj18}. More than two decades after
their discovery, the origin of hot Jupiters (hereafter HJ) remains an outstanding question in
exoplanetary science. Given the conditions so close to the star,
in-situ formation seems unlikely \citep[although see][]{batygin16,boley16}. The two main competing scenarios are disc
migration \citep[e.g.,][]{lin96,kn12} and
high-eccentricity migration \citep[e.g.,][]{wm03,fabrycky07,nagasawa08,wl11,naoz11,
bn12,petrovich15b,anderson16,munoz16}. High-eccentricity migration can be divided into different flavors, one of them being referred to as secular chaos
\citep{wl11}, which is the topic of this paper. In a nutshell,
interactions between three or more planets gradually increase the
eccentricity of the innermost planet until tidal dissipation removes a
large amount of orbital energy to circularize the orbit to a short
period.

Secular chaos and the long-term evolution of planetary systems have a
rich history which finds its roots in the problem of the long-term
stability of our own solar system. Numerical simulations have established
that the solar system is ``marginally'' unstable \citep{laskar09}. The inner solar system (particularly Mercury) may evolve into a
state where catastrophic events happen; the evolution is chaotic and
depends sensitively on the initial conditions, with the timescale
for the onset of chaos comparable to the age of the solar system.
In addition, it is now recognized that the origin of this chaotic evolution
lies in secular resonances. These take place on a timescale much
longer than the orbital period of the planets, and can be studied in
an orbit-averaged framework where the orbital evolution is governed by
the interaction between elliptical and mutually inclined rings.
In the limit of small eccentricities and inclinations,
the evolution of the system can be reduced to a series of
eigenmodes \citep[the Laplace-Lagrange theory; see][]{md99}.
A system of $N$ planets can then be described by $2N$ linear modes,
which depend only on the masses and semi-major axes of the planets. For a test
particle orbiting around the Sun under the perturbations from these
modes, there exists specific locations where the natural precession
frequency of the particle matches one of the
eigenfrequencies. This forms the basis of linear secular resonances.

To understand secular chaos, one needs to go beyond the linear theory.
Motivated by the findings of \citet{laskar08} that the
dynamics of the inner solar system is dominated by secular
effects, \citet{lw11} constructed a non-linear model
of secular interactions to explain Mercury's chaotic dynamics.
The overlap of secular resonances causes Mercury's orbit
to wander chaotically in phase space, gradually increasing its
eccentricity and inclination. In addition to the proximity to
secular resonances, the other key parameter for the onset of
secular chaos is the total angular momentum deficit (AMD)
of a planetary system, i.e., how much angular momentum
would one need to inject into a system to make all the planets
circular and co-planar at the same semi-major axes. AMD
is conserved by secular interactions. \citet{wl11}
found that AMD tends to diffuse chaotically so that the
system evolves towards equipartition between the secular modes.
It is thought that the diffusion of AMD is caused by the
aforementioned overlap of secular resonances, although the
details of this connection remain to be fully understood. The
key finding of \citet{wl11} is that when the AMD
exceeds a critical threshold, its diffusion can increase the
eccentricity and inclination of the inner planet to large
values.

While the work of \citet{wl11} has established secular chaos
as a robust mechanism to achieve near-unity eccentricities
($1-e\lesssim 0.01$) for the inner planet, the precise condition and
the range of timescales for this to happen are unclear.  Moreover,
achieving such high eccentricities is only part of the hot Jupiters
formation story. Indeed, one still needs a mechanism to circularize
the planet's orbit and to bring it from semi-major axis of several
au's to $\sim 0.05$ au. So far, most studies of high-e migration (see
references above), with the exception of \citet{wu18} and \citet{vla18} have relied on the parametrized weak-friction
theory of equilibrium tides \citep{alexander73,hut81} or ad hoc tidal
constant-$Q$ models. In order achieve tidal circularization within 5
Gyrs in high-e migration scenarios, the giant planet should be more
dissipative than Jupiter by more than a factor of 10 \citep{socrates12,anderson16}. It has been recognized the tidal
response of a fluid body (such as a giant planet) is a complicated
problem that involves the excitation of internal waves and modes at
different frequencies \citep[see, e.g.,][]{ogilvie14}. \citet{mardling95a,mardling95b} and \citet{ivanov04,ivanov07} used a more complex 
description of the tidal response, called dynamical tides,
to show that the excitation of internal modes (mainly the
f-mode) can lead to very efficient deposition of orbital energy 
into fluid motion. This mechanism has been recently
put forward by \citet{vl18}, \citet{wu18} and \citet{vla18} to allow
for enhanced tidal dissipation of giant planets on high-eccentricity orbits.

Secular chaos is a process involving at least three planets, and takes place on millions to billions of years. In addition, the very high eccentricities that can potentially be achieved require a very fine accuracy to properly resolve pericenter passages. This makes direct $N$-body integrations impractical to survey the large parameter space covered by a system of three planets. The work by \citet{lw14} represents so far the only attempt at exploring the outcome of secular chaos, using at limited set of 100 simulations. 
In addition, tidal disruption of the planet, which severely limits the efficiency of Lidov-Kozai migration of giant planets \citep{petrovich15a,anderson16,munoz16}, has not been systematically explored in the secular chaos scenario. Furthermore, to predict the final stellar obliquities of HJ systems, one must include the effect of spin-orbit coupling, as the variation of the stellar spin during the high-$e$ migration plays an important role in determining the spin-orbit evolution \citep{storch14,storch15,anderson16,storch17}.

In this paper, we perform the first systematic study of HJ formation via the
secular chaos mechanism.  Our work goes beyond that of \citet{wl11} and \citet{lw14} in several ways: (i) We use a secular Gaussian ring code \citep{md99,ttk09} to compute the long-term evolution of 3-planet systems and explore the conditions for the onset of secular chaos; (ii) Using the ring code we carry out a large suite of calculations (population synthesis) for various initial conditions and parameters; (iii) We incorporate dynamical tides \citep[using the method developed in][]{vla18}, which can significantly affect the outcomes of high-$e$ migration; (iv) We include all relevant physical effects (tidal disruption, short-range forces, spin-orbit coupling and magnetic braking of host stars) to determine the properties of HJs formed in the secular chaos scenario.

This paper is organized as follows. Section \ref{sec:sechaos} contains a discussion of the key features of secular chaos that are relevant to our study. Section \ref{sec:methods} presents our numerical methods and the implementation of short-range forces, spin evolution and tidal interactions. A first example of secular chaos is studied in Section \ref{sec:ill}, while a more thorough exploration (for large parameter space and initial conditions) is presented in Section \ref{sec:sims}. A discussion of these results and concluding remarks can be found in Sections \ref{sec:discussion} and \ref{sec:conclu}.

\section{Secular chaos: key features}
\label{sec:sechaos}

\subsection{Secular diffusion of AMD and the onset of chaos}
\label{sec:amd}
The total angular momentum deficit (AMD) of a system of $N$ planets orbiting a star of mass $M_*$ is defined as:
\begin{equation}
\label{eq:amd}
A=\sum_{k=1}^N \Lambda_k \left(1-\sqrt{1-e_k^2}\cos i_k\right),
\end{equation}
where 
\begin{equation}
\label{eq:lambda}
\Lambda_k=\frac{m_k M_*}{m_k+M_*}\sqrt{G(m_k+M_*)a_k}
\end{equation}
is the circular angular momentum of planet $k$ \citep[with mass $m_k$ and standard notations for the orbital elements, see e.g.,][]{md99}. Because secular perturbations conserve the total angular momentm of the system and the individual orbital energy of each planet, AMD is conserved by secular interactions. As noted by \citet{laskar97}, although the total AMD is conserved, the individual AMD of each planet can evolve significantly, leading to an overall diffusion of AMD on very long timescales. In fact, \citet{wl11} have noted that the AMD resembles the free energy of a system, and will tend to equipartition between the secular modes that govern the long term evolution of planetary systems (in the absence of significant mean-motion resonances). In practice, this means that a large fraction of the total AMD can be eventually deposited into one planet, raising its eccentricity and inclination to high values. If the planet reaches a pericenter distance small enough for tidal effects to become significant, a hot Jupiter can be formed. Therefore, the initial AMD of the system (which is entirely set by the initial conditions) is crucial in determining the final fate of the system, as systems with more AMD are more prone to this chaotic diffusion. \citet{wl11} noted that if all the AMD of the system could be deposited into the innermost orbit, bringing the inner planet to a pericenter $p_{\rm min}$ requires that the system have a minimum amount of AMD given by
\begin{equation}
\label{eq:amdcrit}
\text{A}_{\rm c} = \Lambda_1 \left[ 1 - 0.3\left(\frac{p_{\rm min}}{0.05~{\rm au}}\right)^{1/2}\left(\frac{a_1}{1~{\rm au}}\right)^{-1/2}\cos i_{1,f} \right],
\end{equation}
where $i_{1,f}$ is the final inclination of the inner planet, and we have assumed $p_{\rm min}\ll a_1$. \citet{wl11} pointed out that well-spaced systems of 3 giant planets with moderate eccentricities and inclinations will easily satisfy such requirement, and are therefore prone to chaotic diffusion.

We remark that when the AMD of the system has chaotically diffused to the point where it is predominantly in the inner planet, the conservation of AMD implies $\sqrt{1-e_1^2}\cos i_1\approx$ constant. This implies that at very large eccentricities, the dynamics of the inner planet shares some similarities with the Lidov-Kozai mechanism. 

\subsection{Proximity to secular resonances}
\label{sec:secres}
Having a large AMD is a necessary but not sufficient condition for chaotic diffusion to happen. For instance in a planetary system consisting of Jupiter and Saturn around the Sun, the presence of an equal-mass, coplanar binary companion at 1000 au with an eccentricity of 0.1 would provide the system with a total AMD far in excess of the minimum amount required by the analysis of the previous section, and yet one can numerically verify that the system would remain stable for more than 10 Gyr. 

It is useful to consider a simple model based on the linear theory of secular interactions \citep[see, e.g.,][]{md99}. A system of $N$ planets will have $2N-1$ non-zero secular frequencies ($N$ related to apsidal precession and $N-1$ related to nodal precession). Linear secular resonances can happen when the natural precession frequency of one of the planets is close to one of these secular forcing frequencies.
Using a simple model of a test particle perturbed by two of such frequencies, \citet{lw11} showed that secular resonances would cause the orbit of the test particle to become chaotic. 

Consider a toy model of a three-planet system, in which the innermost planet (labelled 1) is treated as a test-particle and is perturbed by two outer planets (labelled $k=\{2,3\}$). The precession induced by planet $k$ on planet 1 is \citep[assuming $a_k\gg a_1$; see e.g. eq. 18 of][]{liu15},
\begin{equation}
\dot{\omega}_{1k}=\frac{3}{4}(1-e_1^2)^{1/2}t_{1k}^{-1},
\end{equation}
where $t_{1k}$ is the quadrupole timescale given by
\begin{equation}
t_{1k}=\frac{M_*+m_1}{m_k}\left(\frac{a_k}{a_1}\right)^3(1-e_k^2)^{3/2} n_1^{-1},
\end{equation}
and with $n_1=\sqrt{G(\Ms+m_1)a_1^{-3}}$ the mean motion of planet 1.
Adding the effect of short-range forces (labelled SRF, see equations \ref{eq:gr}--\ref{eq:rot} in Section \ref{sec:srf}) we get the total precession rate of planet 1:
\begin{equation}
\label{eq:w1}
\dot{\omega}_1=\dot{\omega}_{12}+\dot{\omega}_{13}+\dot{\omega}_{SRF}.
\end{equation}
If we further assume that the system is coplanar, the outer two planets are governed by two modes, whose frequencies are $g_\alpha$ and $g_\beta$.
A secular resonance will happen when $\dot{\omega}_1\approx g_{\alpha,\beta}$. The expression for $g_{\alpha,\beta}$ is
\begin{equation}
\label{eq:gab}
g_{\alpha,\beta}=\frac{A_1(\Lambda_2+\Lambda_3) \pm \sqrt{A_1^2(\Lambda_2-\Lambda_3)^2+4A_2^2\Lambda_2\Lambda_3}}{2\Lambda_2 \Lambda_3},
\end{equation}
where
\begin{equation}
A_k=\frac{1}{4} G m_2 m_3~\frac{a_2^2}{a_3^3}~b_{3/2}^{(k)}(a_2/a_3).
\end{equation}
and $b_{3/2}^{(k)}$ are the Laplace coefficients \citep[see, e.g.,][]{md99}. For completeness we note that a system of two planets has one non-zero frequency associated with nodal precession when the mutual inclination is non-zero. This precession is retrograde, at a rate given by $-(g_\alpha+g_\beta)$

\begin{figure}
    \begin{center}
    \includegraphics[scale=0.85]{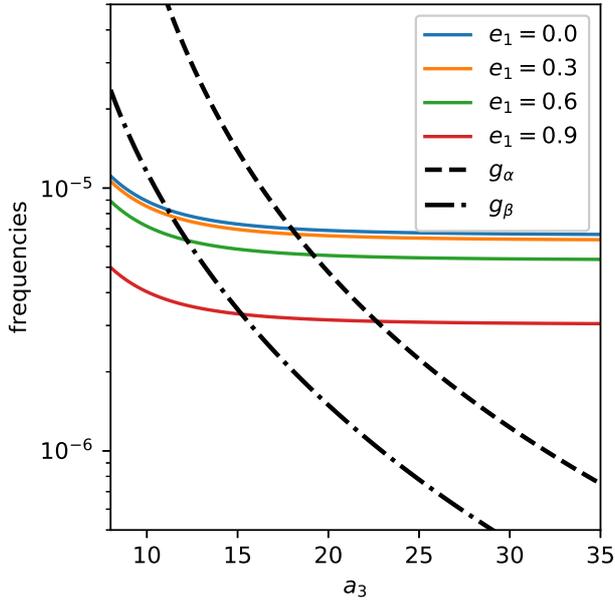}
    \caption{Various secular frequencies as a function of $a_3$ for the initial conditions of Table \ref{tab:ICfidu}. The black thick dashed curves are the two linear frequencies associated with planets 2 and 3, assuming planet 1 is a test particle. The coloured curves are the natural frequency of planet 1, for different eccentricities (see equation \ref{eq:w1}). Secular resonances occur when a dashed curve meets a solid curve. Hence resonances with the $g_\alpha$ mode occur for $a_3$ between 18 and 23 au, while resonances with the $g_\beta$ mode occur for $a_3$ between 11 and 15 au.
    } 
    \label{fig:a3freq}
    \end{center}
\end{figure}	
Although we assume that the dynamics of the outer two planets is dominated by their linear frequencies, we consider arbitrary eccentricities for the inner planet.
Thus, the location of the secular resonance changes as the eccentricity of the inner planet evolves. \citet{lw11} also found that including inclinations broadened the region of parameter space that is available for chaotic evolution. In Sections \ref{sec:ill} and \ref{sec:sims} we will show numerical evidence that a system close to a secular resonance is a necessary condition for the onset of secular chaos, along with having a large reservoir of AMD. 

Figure \ref{fig:a3freq} shows the precession rate of the inner test particle (eq. \ref{eq:w1}) for different eccentricities, as well as the frequencies $g_{\alpha,\beta}$ (eq. \ref{eq:gab}). In anticipation of Section \ref{sec:a3var} where we run numerical experiments varying $a_3$ only as a control parameter, we have plotted these frequencies keeping everything fixed but $a_3$. Here the innermost planet is a test-mass at $a_1=1~{\rm au}$, the outer two planets have the same initial conditions as in Section \ref{sec:a3var}: $a_2=6~{\rm au}$, $m_2=1.9~\Mj$ and $m_3=2.8~\Mj$. Figure \ref{fig:a3freq} illustrates how, with this set of initial conditions, one can expect secular resonances between the $g_\alpha$ mode and the free precession rate of the test particle to occur for $a_3$ between 18 and 23 au, while resonances with the $g_\beta$ mode occur for $a_3$ between 11 and 15 au. These resonances will excite the eccentricity of the test particle to large values. This analysis will prove useful when studying the dynamics of three planets in Section \ref{sec:ill}.

\section{Simulating high-e migration via Secular Chaos: Methods}
\label{sec:methods}

\subsection{Numerical methods}
\label{sec:num}

Numerically simulating systems of three planets undergoing secular chaos is challenging because the chaotic diffusion timescale is usually many orders of magnitude longer than the orbital period of the inner planet. This makes carrying direct $N$-body simulations for a large ensemble of systems an expensive task. On the other hand, the much faster secular (i.e. orbit-averaged) approximations usually employed in the literature cannot be applied here. The hierarchical approximation (most famously used to explore Lidov-Kozai oscillations) requires  a large orbital separation between various bodies, and is rather cumbersome   to apply to four-body systems \citep{hamers15}. Also, the Laplace-Lagrange approximation is based on expansions in powers of the eccentricities and inclinations, and therefore does not apply to systems where the eccentricities can reach near-unity values and the inclinations become larger than $\pi/2$. 

Luckily a third alternative exists, which is due to Gauss \citep{md99}. It is also an orbit-averaged method, where the interaction between two planets reduces to the interaction between two rings. Gauss's aim was to compute as accurately as possible the force exerted by one ring on the other. An algorithm to implement this method was given by \cite{ttk09}, and a code was made publicly available by Will M. Farr \footnote{https://github.com/farr/Rings}. We have modified this \textsc{rings} code to include short-range forces and tides, as described in Sections \ref{sec:srf}, \ref{sec:j2} and \ref{sec:tides}. It is important to note that when dynamical tides are switched on, the timestep of the integrator is reduced to the orbital period of the innermost planet. Despite this, the \textsc{rings} code is still more efficient than direct $N$-body simulations.

Each of our simulations was integrated for a maximum of 2 Gyr. The integration are stopped earlier in the following cases:
\begin{itemize}
\item A hot Jupiter (here defined as a planet with $a<0.08~{\rm au}$ and $e<0.005$) has formed.
\item The planet's pericenter comes within $2\rt$, where
\begin{equation}
\label{eq:rt}
\rt=\Rp\left(\frac{\Ms}{\Mp}\right)^{1/3}
\end{equation}
is the tidal radius (see Section \ref{sec:rt} for more details). We assume this causes complete disruption of the planet.
\item A pair of planets crosses the stability criterion defined in Section \ref{sec:stab}
\end{itemize}
We set the integration time to be 2 Gyr because when a system evolves chaotically, we cannot achieve a good conservation of AMD over timescales longer than 2 to 3 Gyr.

\subsection{Short-range forces}
\label{sec:srf}

When the inner planet (mass $\Mp$ and radius $\Rp$) approaches the host star on a high-eccentricity orbit, its orbital evolution can be strongly influenced by short-range forces (SRFs). The most important SRFs arise from the first-order relativistic correction (labeled GR), the static tidal bulge of the planet (Tides), and the rotational distortion of the planet (Rot). 
All these forces cause the eccentricity vector of the planet to precess in a prograde manner, and this procession can limit the maximum eccentricity that a planet can reach  \citep[see, e.g.,][]{wm03,liu15}. Under the combined effect of these forces, the the eccentricity vector $\bm{e}$ of the inner planet precesses at a rate given by:
\begin{equation}
\dd{\bm{e}}{t}=\left(\dot{\omega}_{\rm GR} +\dot{\omega}_{\rm Tides}+\dot{\omega}_{\rm Rot}\right)\Lhat\times \bm{e}
\end{equation}
where $\Lhat$ is the unit angular momentum vector of the orbit, and the precession frequencies are given by:
\begin{align}
\label{eq:gr}&\dot{\omega}_{\rm GR} = \frac{3 G(\Ms+\Mp)}{c^2 a(1-e^2)}n,\\
\label{eq:tid}&\dot{\omega}_{\rm Tides} = \frac{15}{2}\ktwop\frac{\Ms}{\Mp}\left(\frac{\Rp}{a}\right)^5 \frac{f_4(e)}{j^{10}}n,\\
\label{eq:rot}&\dot{\omega}_{\rm Rot} =\frac{3}{2}k_{qp}\left(\frac{\Rp}{a}\right)^2\frac{\hat{\Omega}_{\rm p}^2}{j^4}n.
\end{align}
We have introduced the speed of light in vacuum $c$, the tidal Love number of the planet $\ktwop$, a rotational distortion coefficient $k_{qp}$, and $\hat{\Omega}_p$ the planet rotation rate in units of the breakup frequency, $\hat{\Omega}_{\rm p}=\Op(G\Mp \Rp^{-3})^{-1/2}$. We have also introduced $j=\sqrt{1-e^2}$, and the function $f_4$ is defined in Appendix \ref{app:wft}, as well as all the functions $f_1$ to $f_5$ that will appear throughout this paper. Fiducial values for the physical parameters used in the implementation of SRFs and equilibrium tide dissipation are given in Table \ref{tab:phys}.

\setlength{\tabcolsep}{4pt}
\begin{table}
  \caption{Notations and values of various parameters.}
  \label{tab:phys}
  \begin{tabular}{lc}
    \hline\hline
    Quantity &  fiducial value\\
    \hline
	Stellar moment of inertia constant $\ks$ & 0.1 \\
	Planet moment of inertia constant $\kp$ & 0.25\\
	Stellar rotational distortion coefficient $k_{q*}$ & 0.05\\
	Planet rotational distortion coefficient $k_{qp}$ & 0.17\\
	Planet tidal Love number $k_{\rm 2p}$ & 0.37 \\
	Tidal time lag $\Delta t_{\rm L}$ & 1\,s \\
	\hline
  \end{tabular}
\end{table}

\subsection{Stellar spin dynamics}
\label{sec:j2}
The orientation of the stellar spin with respect to the planet's orbit (the spin-orbit misalignment) provides an important diagnostic for the formation history of hot Jupiters. While most efforts have been directed at generating large planetary inclinations (e.g. through the Lidov-Kozai mechanism), it has only been recently appreciated that the spin of the star can also undergo dramatic evolution, and even become chaotic \citep{storch14,storch15,anderson16}. The star has a spin rate $\Os$, moment of inertia constant $\ks$ and radius $\Rs$. Its spin angular momentum $\Ssv$ is given by $\Ssv=\Is\Os\Sshat$ with $\Is=\ks\Ms\Rs^2$. The possible chaotic spin evolution can lead to large spin-orbit angles $\theta_{\rm sl}$ (which we define here through $\costsl=\Lhat\cdot\Sshat$). The evolution of the stellar spin is given by
\begin{equation}
\dd{\Ssv}{t}=\Omega_{\rm ps}\Lhat\times\Ssv-\alpha_{\rm MB}\Is\Os^3\Sshat.
\end{equation}
The first term describes the precession of the stellar spin $\Ssv$ around the planet's angular momentum $\bm{L}$, while the second term describes the stellar spin-down due to magnetic braking, with an efficiency parametrized by $\alpha_{MB}=1.5\times 10^{-14}$yr for Sun-like stars \citep{bo09}. The precession rate is given by
\begin{equation}
\Omega_{\rm ps}=-\frac{3}{2}\frac{k_{q*}}{\ks}\frac{\Mp}{\Ms}\left(\frac{\Rs}{a}\right)^3\frac{\Os}{j^3}\costsl,
\end{equation}
where $k_{q*}$ is a rotational distortion coefficient related to the usual $J_2$ parameter of the star by $J_2=k_{q*}\hat{\Omega}_*^2$ with $\hat{\Omega}_*$ the star rotation rate in units of  the breakup frequency, $\hat{\Omega}_*=\Os(G\Ms \Rs^{-3})^{-1/2}$.

The back-reaction of the stellar quadrupole on the planet's orbit takes the form
\begin{align}
&\dd{\bm{L}}{t}=\Omega_{\rm ps}\Ssv\times\Lhat,\\
&\dd{\bm{e}}{t}=-\dot{\omega}_*\left[\costsl\Sshat\times\bm{e}+\frac{1}{2}(1-5\cos^2\theta_{\rm sl})\Lhat\times\bm{e}\right],
\end{align}
where $\dot{\omega}_*$ is the characteristic precession rate of  the planet's:
\begin{equation}
\dot{\omega}_*=\frac{3}{2}k_{q*}\left(\frac{\Rp}{a}\right)^2\frac{\hat{\Omega}_*}{j^4}n.
\end{equation}

\subsection{Tidal interactions}
\label{sec:tides}

\subsubsection{Weak friction theory}
\label{sec:wft}
As noted in Section \ref{sec:intro}, most previous works on high-eccentricity migration have adopted the weak friction model of static tides.
In this idealized model, tidal dissipation is parametrized by a constant time lag $\dtl$ and the internal structure of the planet is represented by the tidal Love number $\ktwop$. It is useful to introduce the timescale $t_a$ given by \citep[e.g.,][]{correia11,anderson16}:
\begin{equation}
\label{eq:ta}
t_a^{-1}=6\ktwop\dtl\frac{\Ms}{\Mp}\left(\frac{\Rp}{a}\right)^5 n^2.
\end{equation}
Let $\Spv=\Ip\Op\Sphat$ be the spin angular momentum of the planet, with $\Ip=\kp\Mp\Rp^2$ being the planet's moment of inertia, $\kp$ a coefficient describing the mass distribution inside the planet, and $\Op$ the rotation rate. Assuming that the planet's spin remains aligned with the orbital angular momentum ($\Sphat=\Lhat$), the spin evolution is governed by
\begin{equation}
\dd{\Sp}{t}=-\frac{L}{2t_a j^{13}}\left[j^3f_5(e)\frac{\Op}{n}-f_2(e)\right],
\end{equation}
The change in the orbit is then given by
\begin{align}
\dd{\bm{L}}{t}&=-\dd{\Sp}{t}\Lhat ,\\
\dd{\bm{e}}{t}&=\frac{1}{2t_a j^{13}}\left[\frac{11}{2}j^3f_4(e)\frac{\Op}{n}-9f_3(e) \right]\bm{e}.
\end{align}

\subsubsection{Dynamical tides}
\label{sec:dt}
Tidal forcing from the star can excite internal oscillations inside the planet. This fluid motion can be decomposed into different oscillatory modes. For highly eccentric orbits, the excitation of these modes occurs around pericentre passages, which last a very short time compared to the orbital period. To this end, \citet{ivanov04} used an impulse approximation to model the mode energy evolution: once per orbit, at each pericenter passage, the mode recieves a kick in energy. This energy transfer between the mode and the orbit will in turn change the orbit of the planet. This process was written down by \citet{ivanov04} in the form of an iterative map. \citet{vl18} later refined this map to include mode damping, and showed that depending on the mode property of the planet and the pericenter distance and eccentricity of the orbit, the mode evolution can exhibit several distinct behaviors: low-amplitude quasi-steady oscillation, resonant oscillations and chaotic growth \citep[see also][]{mardling96,wu18}. \citet{wu18} and \citet{vla18} have recently examined the effect of chaotic tides in the formation of HJs via Lidov-Kozai oscillations in binary star systems. Following their work, we focus on the fundamental mode (or f-mode) as it has the largest fluid response, and therefore generates the most dissipation. In particular, \cite{vl18} found that inertial modes are less important for dissipating energy than f-modes when using an $\gamma=2$ polytropic model for the planet.

Let us now consider highly eccentric planetary orbits. At each pericenter passage, energy is transferred between the mode and the orbit. It is the phase of the mode that determines the direction of energy transfer \citep[from the orbit to the mode or the other way around, see, e.g.,][]{kochanek92}. 
Therefore to determine the effect of dynamical tides, it is necessary to follow the complex mode amplitude on the orbit-to-orbit basis. This is challenging because high-eccentricity migration operates on a secular timescale and involves a large number of orbits. This is particularly the case for secular chaos, since it can take $10^9$ orbits for the inner planet to be pushed to a sufficiently high eccentricity. Here we follow the strategy developed by \citet{vla18}. We summarize the key steps below.
\\

\noindent\textbf{Switching on dynamical tides.}  
Consider a mode with frequency $\sigma$ whose complex amplitude is $c_{k-1}$ just before the $k$-th pericenter passage. The new mode amplitude just before the $(k+1)$-th passage is \citep[see][]{vl18}:
\begin{equation}
c_k=(c_{k-1}+\Delta c){\rm e}^{-{\rm i}\sigma P_k}.
\end{equation}
Hence at each pericenter passage, the mode amplitude changes by $\Delta c$ (a real quantity which is calculated using the current orbital parameters), and the orbital period changes to $P_k$. When the mode has no initial energy, the quantity that determines the evolution of the system is $\Delta\hat{P}\equiv\sigma\Delta P$. Here $\Delta P=P_1-P_0$ is the change in orbital period due to the energy transfer between the mode and the orbit, $\Delta E$ (with $\Delta c\propto\sqrt{\Delta E}$). When $\Delta\hat{P}\gtrsim 1$, the change in mode phase between orbits is nearly random (mod $2\pi$), and the mode energy grows chaotically \citep[see][for the behaviour of the system as a function of $\Delta\hat{P}$.]{vl18}. Furthermore,  \citet{wu18} and \citet{vla18} have refine this model to account for the fact that the mode can contain a pre-existing amount of energy $E_{k-1}$ just before the $k$-th passage. This gives a modified $\Delta\hat{P}_k(E_{k-1})$ that depends on $E_{k-1}$  \citep[see Eq. 38 of ][]{vla18}. 
The critical value of $\Delta \hat{P}_k(E_{k-1})$ necessary for chaotic mode growth depends on the orbital period and is smallest when the mode frequency is in resonance with the orbital frequency \cite[see Fig. 4 of][]{vla18}. In practice, we begin to track the mode evolution when $\Delta \hat{P}_k(E_{k-1}) >0.01$ to ensure that the model captures the onset of chaotic behaviour. 
At this point, the integration timestep is reduced to the orbital period and the orbital parameters are updated at each timestep to account for the effect of dynamical tides.\\

\noindent\textbf{Evolving the orbit.}
Just before the $k$-th pericenter passage, the planet has a semi-major axis $a_{k-1}$, eccentricity $e_{k-1}$ and mode amplitude $c_{k-1}$. To compute the orbital parameters after the $k$-th pericenter passage we follow these steps:
\begin{enumerate}[leftmargin=\parindent,align=left,labelwidth=\parindent,labelsep=0pt]
\item Calculate $\Delta E_k$, the energy transfer to the mode when the mode has zero amplitude prior to the pericenter passage. Details can be found in Appendix \ref{app:dt}. Obtain the (normalized) change in mode amplitude $\Delta \tilde{c}=\sqrt{\Delta E_k/|E_{B,0}|}$, where $E_{B,0}$ is the initial orbital energy.
\item Calculate the energy transfer during the $k-$th passage: $\Delta \tilde{E}_k=|\tilde{c}_{k-1}+\Delta \tilde{c}|^2-|\tilde{c}_{k-1}|^2$, with $\Delta \tilde{E}_k = \Delta E_k/|E_{B,0}|$. Obtain the new orbital energy $E_{B,k}=E_{B,k-1}-\Delta E_k$.
\item Obtain the new semi-major axis and eccentricity (assuming conservation of orbital angular momentum during the pericenter passage):
\begin{align}
a_k &= \frac{E_{B,k-1}}{E_{B,k}}a_{k-1} \\
e_k &= \left[1-\frac{E_{B,k}}{E_{B,k-1}}(1-e_{k-1}^2) \right]^{1/2}. 
\end{align}
\item Evolve the complex mode amplitude over one orbit to obtain its value just before the $(k+1)$-th passage: $c_{k}=(c_{k-1}+\Delta c){\rm e}^{-{\rm i}\sigma_k P_k}$. The period $P_k$ follows directly from $a_k$. In addition, the mode frequency $\sigma_k$ changes as the system evolves because it depends on the planet's spin (see Appendix \ref{app:dt}).
\end{enumerate}

\noindent\textbf{Non-linear dissipation.}
The procedure outlined above will deposit more and more energy in the mode. After many orbits the mode energy can become so large that it will come dangerously close to the planet's binding energy $\sim G \Mp^2/\Rp$. We therefore assume that when the mode energy reaches a maximum of $E_{\rm max}=0.1 G \Mp^2/\Rp$, non-linear effects will dissipate the energy very efficiently over one orbital period, leaving it at some residual energy $E_{\rm resid}=0.01 E_{\rm max}$. At this point, we rescale the mode amplitude  so that $|\tilde{c}_k|^2=E_{\rm resid}/|E_{B,0}|$, and the mode energy can start growing again, as outlined above. See \citet{vl18} for a discussion regarding the choice and influence of the values of $E_{\rm resid}$ and $E_{\rm max}$. \\

\noindent\textbf{Switching off dynamical tides.}
This occurs when $|\Delta \hat{P}_k (E_{k-1}| < \Delta \hat{P}_{\rm crit} \sim 1$ while the orbital eccentricity is still large. Typically, this condition is met just after the mode energy has dissipated from $E_{\rm max}$ to $E_{\rm resid}$ and the system satisfies $\Delta \hat{P}_{\rm crit} \lesssim 1$ \citep[see Equation 51 of][]{vla18}.
It is then no longer necessary to use the orbital period as the integration timestep, and one can resume using a much longer timestep.

\subsection{Tidal disruption and maximum eccentricity}
\label{sec:rt}

Of major concern for high-eccentricity migration mechanisms is the possibility that the planet be tidally disrupted when approaching the star too closely. 
We parametrize the critical tidal disruption radius by $\eta \rt$, with $\rt$ given by eq. (\ref{eq:rt}) and $\eta$ of order a few.

Simulations of the Lidov-Kozai formation of hot Jupiter in binary stars by \citet{naoz12} have used $\eta=1.67$. Based on the SPH simulations of tidal disruption by \citet{faber05} many authors have used $\eta=2.16$. In particular, \citet{wl11} chose this value for their work on secular chaos. More recent simulations by \citet{guillochon11} suggested $\eta=2.7$. Using this value, \citet{petrovich15a}, \citet{anderson16} and \cite{munoz16} found that a large fraction of planets undergoing Lidov-Kozai oscillations in binary stars would be disrupted, while they could have survived using a lower value of $\eta$. As discussed in Section \ref{sec:num}, we stop our integrations when the pericenter distance reaches $2\rt$, in order to be able to analyse our results as a function of $\eta$.

\subsection{Stability of planetary systems}
\label{sec:stab}
The process of AMD diffusion and the secular approach employed in this paper conserve the energy of each orbit by construction, and therefore the planetary semi-major axis do not change over time. One concern for this secular approach is whether all the systems we simulate in this paper would remain stable if the semi-major axis were allowed to change under the effect of planet-planet interactions. While there exists no analytic criterion for the stability of 2-planet systems with arbitrary eccentricities and inclinations (let alone for more than 2 planets), numerous papers have investigated this issue numerically. 

For the mass and semi-major axis regime we are interested in, the most relevant work is that of \citet{petrovich15c}. Analysing a large ensemble of simulations, he proposed that 2-planet systems would be stable (resp. unstable) if $\Cp>1$ (resp. $\Cp<1$), where $\Cp$ is defined as:
\begin{equation}
\label{eq:stabP15}
\Cp=\frac{p_{k+1}/q_k}{2.4\max\left(\frac{m_{k}}{M_*},\frac{m_{k+1}}{M_*}\right)^{1/3}\left(\frac{a_{k+1}}{a_{k}} \right)^{1/2}+1.15},
\end{equation}
with $p=a(1-e)$ the pericenter distance and $q =a(1+e)$ the apocenter distance.
For very low mass planets, this criterion means that the two planets become unstable when the ratio of the pericenter $p_{k+1}$ of the outer planet to the apocenter $q_k$ of the inner one is less than 1.15. The additional term in the denominator is a correction due to the finite planetary masses and separations. \citet{petrovich15c} found that for mutual inclinations lower than 40 degrees (i.e., not in the Lidov-Kozai regime), there was no need to include an inclination dependence into the criterion. Similarly, taking into account the relative orientation of the ellipses did not significantly improve the robustness of the criterion. 

It is interesting to note that for the fiducial example given in Table 1 of \citet{wl11}, the outer two planets would be marginally unstable according to this criterion ($\Cp=0.94$), and yet remained stable throughout the $3\times 10^{8}$ years of the simulation. 
We see three reason for why the example of \citet{wl11} did not suffer from dynamical instabilities. First, Petrovich's criterion was a single fit to a large ensemble of simulations, and therefore must be interpreted in a statistical sense. The system shown in \citet{wl11} is only marginally unstable under this criterion, and nothing guaranties that it will go unstable for the duration of the integration. This brings us to the second point, which is that Petrovich's criterion was the result of numerical integrations up to $10^{8}$ orbits of the inner planet, whereas this particular simulation ran for about $20\times10^6$ inner orbital periods. Finally, and perhaps the most intriguing point, the presence of an inner third planet might interfere with the evolution of the outer two planets, in a way to stabilize them, by absorbing any excess of AMD and therefore preventing the outer two planets from coming too close to each other. The fact that AMD diffusion can help increase the overall stability of a planetary system by getting rid of its ``weakest'' members was already hypothesized by \citet{lw14}. If this hypothesis is correct, it would have interesting consequences on the stability of planetary systems, and would deserve a careful study of its own.

Yet another point to note is that the Petrovich criterion applies to the initial conditions of the systems, while we are seeking a more ``instantaneous'' criterion to determine when the secular integration should be stopped. 
For all the reasons given above, it is clear that Petrovich's criterion cannot be straightforwardly applied to our study. For the numerical experiments presented in the remainder of this paper, we apply a slightly less constraining criterion: We take the systems to be unstable and stop the secular integration when $\Cp<0.75$. This number is justified using a small suit of $N$-body simulations presented in Appendix \ref{app:stab}.

\section{An illustrative case}
\label{sec:ill}

\subsection{A specific example}
\label{sec:example}

\setlength{\tabcolsep}{4pt}
\begin{table}
  \caption{Initial conditions for three planets.}
  \label{tab:ICfidu}
  \begin{tabular}{lcccccc}
    \hline\hline
    Planet &  mass ($\Mj$) & $a$ (AU) & $e$ & $i$ (deg) & $\omega$ (deg) & $\Omega$ (deg)\\
    \hline
    1 & 1.0 & 1.0 & 0.065 & 4.0 & 60 & 156   \\ 
    2 & 1.9 & 6.0 & 0.18 & 16.0 & 123 & 176  \\ 
    3 & 2.8 & 18.6 & 0.32 & 8.0 &304 & 357 \\ 
    \hline
  \end{tabular}
\end{table}

\begin{figure*}
    \begin{center}
    \includegraphics[scale=0.9]{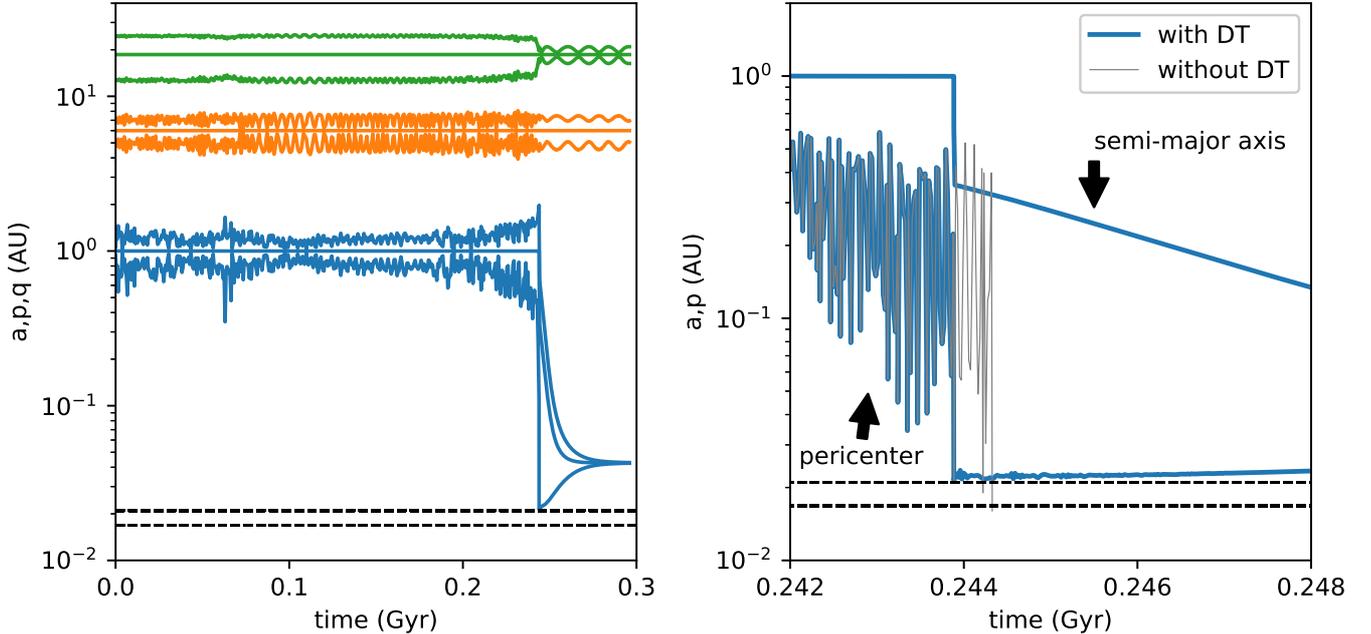}
    \caption{\textit{Left}: Time evolution of planet 1 (blue), 2 (orange) and 3 (green) with initial conditions from Table \ref{tab:ICfidu}. For each planet the lower curve is the pericenter $p$, the middle curve the semi-major axis $a$ and the upper curve the apocenter $q$. The two horizontal dashed lines are the disruption radii at $2.16\rt$ (lower line) and $2.7\rt$ (upper line). The secular evolution keeps the semi-major axis constant, while the eccentricity of the inner planet is gradually excited through secular chaos. After $2.45\times 10^8$ years, the eccentricity of the inner planet reaches 0.978 and dynamical tides quickly dissipate orbital energy. Over the next 50 million years, weak friction tides circularize the orbit while gently bringing the planet closer (final period 3.22 days). A closer view is shown in the \textit{right} panel. The semi-major axis first decays very rapidly because of dynamical tides, and then more gently because of weak friction tides. We also show in grey the pericenter evolution of the same initial system, but integrated with weak friction tides only (no dynamical tides). In this case the planet repeatedly crosses the disruption radii before we stop the integration.}
    \label{fig:DTnoDT}
    \end{center}
\end{figure*}

We start by illustrating secular chaos and the effect of dynamical tides by considering the evolution of a 3-planet system whose initial conditions are given in Table \ref{tab:ICfidu}. The mass of the star is $1\Msun$, its radius $1\Rsun$, and the radius of the inner planet is 1.6 Jupiter radius ($\Rj$).
In Figure \ref{fig:DTnoDT} we show how in this example, secular chaos leads to the orbit of the inner planet to be sufficiently chaotic, such that its eccentricity reaches near-unity. Coupled with our implementation of dynamical tides, this planet eventually becomes a hot Jupiter. In the plot we have marked the lines representing the possible tidal disruption radii at $2.16\rt$ and $2.7\rt$ (see Section \ref{sec:rt}). A closer look shows that the planet avoids disruption as its pericenter never goes below the tidal disruption radius. We have also plotted the evolution of the same system integrated without dynamical tides, in which case the inner orbit crosses the disruption radius and we stop the integration. The effects of dynamical tides are particularly visible on the zoomed-in panel, where we see that they cause the semi-major to decay very rapidly while the eccentricity remains constant ($\sim 0.97$). Subsequent circularization follows, on a timescale set by our weak friction tidal model (see Equation \ref{eq:ta}). 

The end state of this example is a system of three planets including one circularized hot Jupiter with a period of 3.2 days. By diffusing a large amount of AMD to the inner planet, the outer two planets are left in a much more dynamically ``quiet'' state, with their eccentricities experiencing low-amplitude oscillations around 0.2 (planet 2) and 0.08 (planet 3). Therefore planet 3 got rid of most of its AMD. Planet 2 has a comparable eccentricity as it had initially, but the amplitude of its oscillation is much smaller, and no close approach between the outer two planets is possible. Hence by depositing all of its AMD in the inner planet, where it was dissipated through tides, this dynamically ``hot'' system  (i.e. with a large amount of initial AMD) has reached a far more stable state.

\begin{figure}
    \begin{center}
    \includegraphics[scale=0.85]{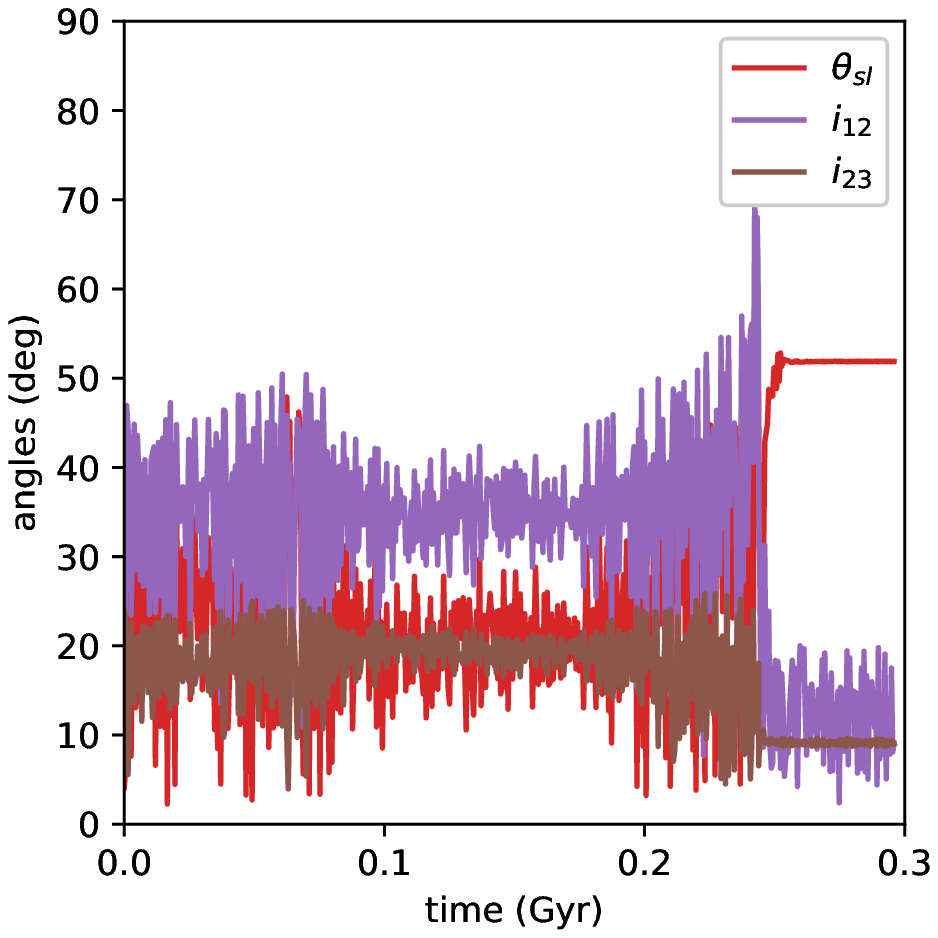}
    \caption{Time evolution of various angles for the system shown in Figure \ref{fig:DTnoDT}: the spin-orbit angle between the star and the inner planet $\theta_{\rm sl}$, and the mutual inclinations between planets 1 and 2 and between 2 and 3, denoted by $i_{12}$ and $i_{23}$, respectively.}
    \label{fig:DT_inc}
    \end{center}
\end{figure}

In Figure \ref{fig:DT_inc} we show $\theta_{\rm sl}$, the angle between the stellar spin axis and inner planet's angular momentum vector. Once the hot Jupiter has formed and is no longer influenced by the outer two planets, the final spin-orbit angle is  $55^{\circ}$. However the mutual inclination angle between planets 1 and 2 (denoted by $i_{12}$) still varies after the hot Jupiter has formed, because planet 2 is still being perturbed by planet 3. This is seen in the the mutual inclination angle between planets 2 and 3 (denoted by $i_{23}$).

Note that our Figure \ref{fig:DTnoDT} follows the same layout as the fiducial example in Figure 1 of \citet{wl11}. Although our initial conditions are slightly different, the qualitative time evolution of the 3-planet system is the same, showing that our orbit-average approach captures all the relevant ingredients of an $N$-body simulation.

\subsection{Varying one control parameter only}
\label{sec:a3var}

We further illustrate the effects of secular chaos and dynamical tides by conducting a series of runs with initial conditions similar to that of Table \ref{tab:ICfidu}. The only difference is that we draw the semi-major axis of the third planet from an uniform distribution between 15 and 31 au. In addition, the longitudes of pericenter and ascending nodes of all three planets are drawn randomly between 0 and 360 degrees. 
We carry two sets of simulations, with and without dynamical tides, and focus on the minimum pericenter reached by the innermost planet in the course of the simulations.

\begin{figure*}
\begin{subfigure}{.5\textwidth}
  \centering
  \includegraphics[width=0.95\linewidth]{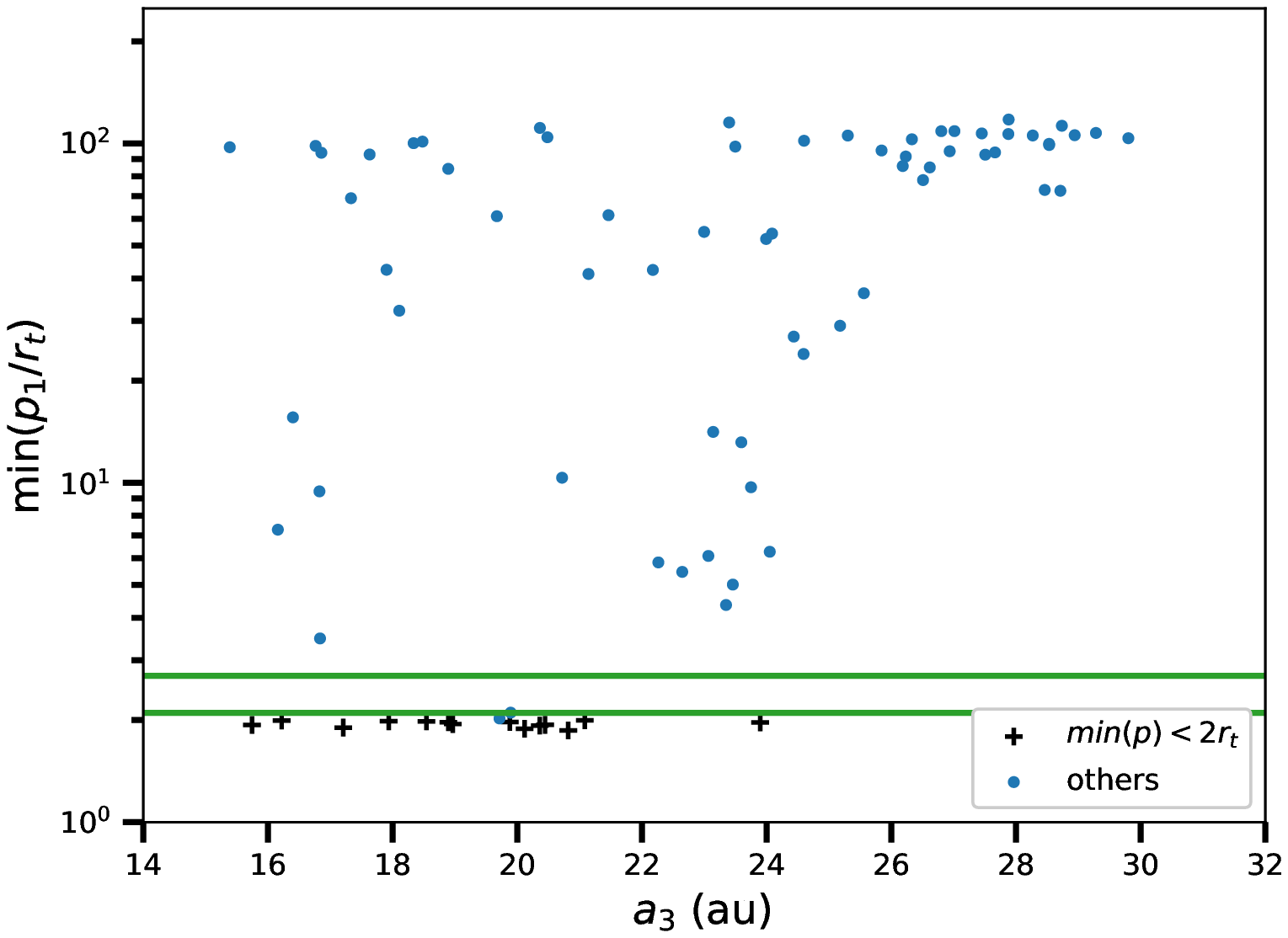}
  \caption{Without dynamical tides}
  \label{fig:sfig1}
\end{subfigure}%
\begin{subfigure}{.5\textwidth}
  \centering
  \includegraphics[width=0.95\linewidth]{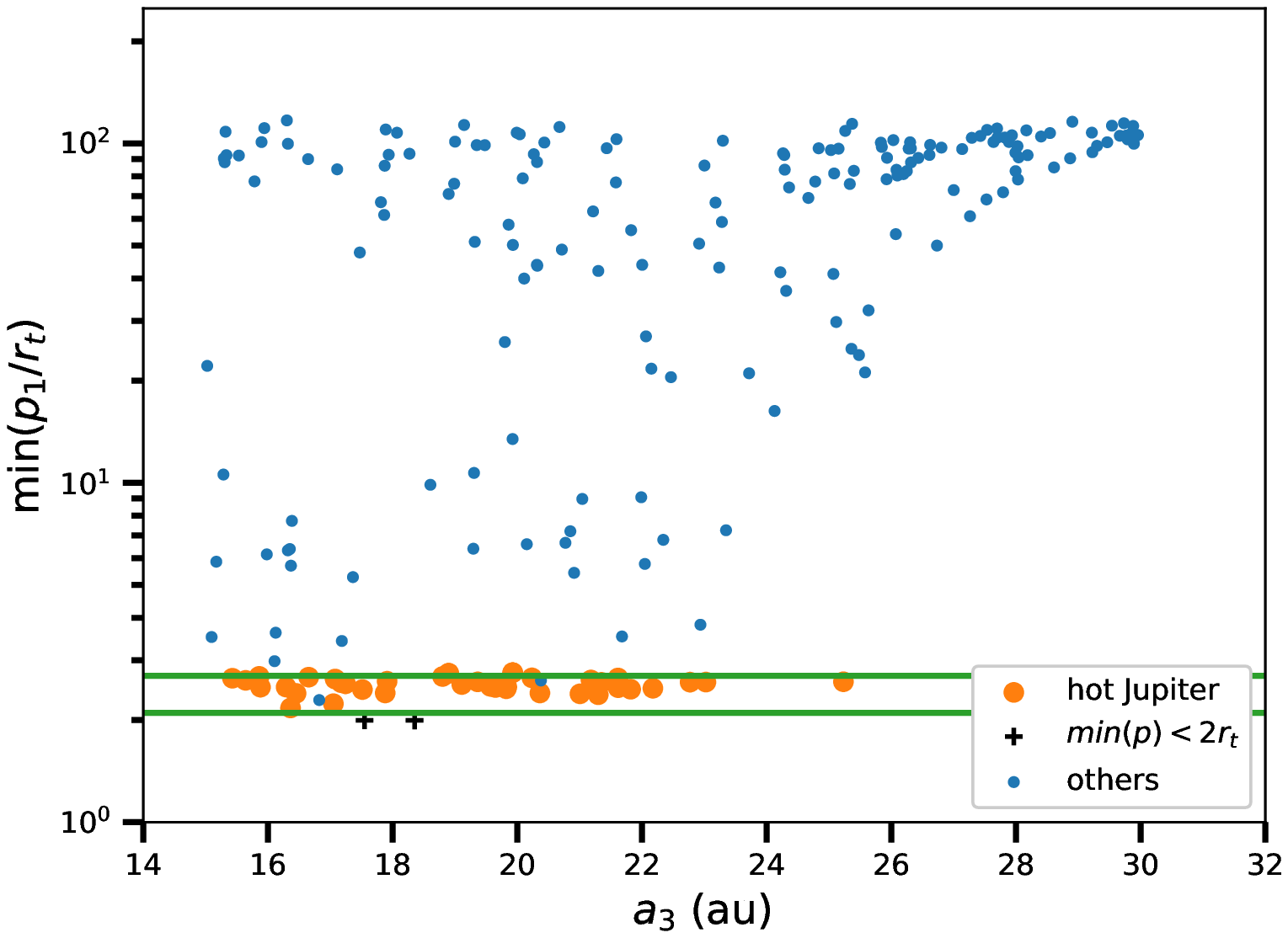}
  \caption{With dynamical tides.}
  \label{fig:sfig2}
\end{subfigure}
\caption{Minimum pericenter distance (in units of the tidal radius, see equation \ref{eq:rt}) of the inner planet without (left) and with (right) dynamical tides, as a function of the semi-major of the outermost planet. Without dynamical tides (only weak friction tides) all the planets that go through an episode of high-eccentricity reach a pericenter lower than $2\rt$ are disrupted (black crosses). With the inclusion of dynamical tides, some hot Jupiters are formed (orange dots). The green horizontal lines indicate the disruption radii at $2.16\rt$ and $2.7\rt$. The initial conditions are those given in Table \ref{tab:ICfidu}, with $a_3$ drawn from a uniform distribution between 15 and 31 au.}
\label{fig:a3pmin}
\end{figure*}

In Figure \ref{fig:a3pmin}a, we show that without dynamical tides, all the planets which go through a phase of very high eccentricity end up with a minimum pericenter lower than $2\rt$ and get disrupted. 

In Figure \ref{fig:a3pmin}b, we show that the inclusion of dynamical tides significantly alters this rather pessimistic conclusion. The minimum pericenter reached by the inner planet is larger than without dynamical tides, possibly preventing planets from being disrupted. Although only a handful of planets became hot Jupiters with a minimum pericenter larger than $2.7\rt$, a significant number of hot Jupiters with minimum pericenter larger than $2.16\rt$ are formed (37 out of 210 simulations). Only two planets reached pericenters below $2\rt$. This numerical experiment illustrates that the choice of disruption radius strongly affects how many hot Jupiters form in a given set of simulations.

The physical reason that dynamical (chaotic) tides save planets from tidal disruption was discussed by \citet{vla18} in the context of Lidov-Kozai migration of giant planets \citep[see also][]{wu18}. In essence, by rapidly reducing the orbital semi-major axis of the inner planet, dynamical tides decouple the inner planet from the influence of the outer planets, thereby preventing it from further eccentricity growth.

Figure \ref{fig:a3pmin} shows that HJs are formed only for $a_3$ between 15.5 au and 23 au. As $a_3$ increases, so does the total amount of AMD of the system, and one might expect that this would in turn lead to more hot Jupiters being formed. This illustrates that the proximity to a secular resonance is key for the onset of secular chaos. As discussed in Section \ref{sec:secres}, we need to compare the precession of the inner planet (eq. \ref{eq:w1}, which depends on the eccentricity and is varying in time) with the two linear frequencies of the outer planets (which vary with $a_3$). Comparing Fig. \ref{fig:a3pmin} with Fig. \ref{fig:a3freq}, we see a clear agreement: a secular resonance occurs when $a_3$ is between 11 au and 23 au. When planet 3 is too distant, both linear frequencies are too slow to resonate with the natural precession frequency of the inner planet, preventing secular chaos from happening. 

\begin{figure}
    \begin{center}
    \includegraphics[scale=0.55]{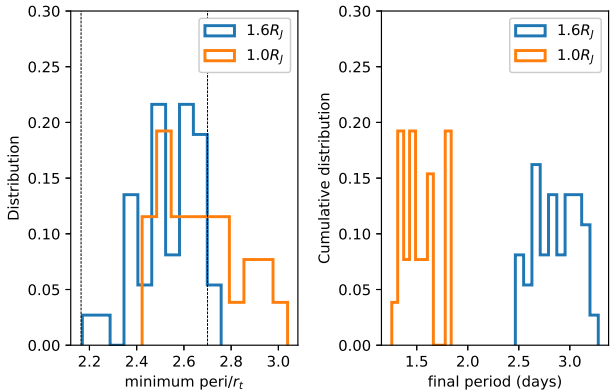}
    \caption{Distribution of the minimum pericenter distance (left panel, in units of tidal radius, see eq. \ref{eq:rt}) and final orbital period (right panel) for all the hot Jupiters formed via secular chaos with dynamical tides. The blue histogram shows hot Jupiters with a radius of $1.6\Rj$ (corresponding to the simulations shown in Figure \ref{fig:a3pmin}b), while the orange histogram shows an additional set of simulations with $\Rp=1\Rj$. On the left panel, the two vertical lines show the disruption radii at 2.16 and 2.7 $\rt$, respectively. Although smaller planets are less prone to disruptions, their final orbital periods are too small to be consistent with observations.} 
    \label{fig:a3_hj_distrib}
    \end{center}
\end{figure}	

In Figure \ref{fig:a3_hj_distrib}, we show the minimum pericenter distance and final period reached by all of the hot Jupiters formed in our simulations with dynamical tides. We also show the results of an additional set of simulations where all the parameters are kept the same apart from the radius of the inner planet ($\Rp=1\Rj$ instead of $\Rp=1.6\Rj$). Such smaller planets are less prone to tidal disruption. However the biggest difference is found when looking at the final period distribution. The planets with a smaller radius ($\Rp=\Rj$) all have periods below 2 days, making them at odds with the observed population of hot Jupiters, whose period peaks between 3 and 4 days. Planets with $\Rp=1.6\Rj$ are more consistent with observations, but still cannot account for the observed systems with periods larger than 3.5 days. This period discrepancy between observations and high-eccentricity migration theories is not restricted to secular chaos, and is also found in the standard Lidov-Kozai migration scenario \citep[see][]{wu07,petrovich15a,anderson16}. Larger period could be produced by \citet{vla18} when coupling Lidov-Kozai oscillations with dynamical tides.

The simulations presented here apply to a specific set of initial conditions, and thus should not necessarily be taken as representative of the outcome of secular chaos. 
In particular, one needs to address whether the high disruption rate (most runs had a minimum pericenter $<2.7\rt$) and short final periods (most runs form a hot Jupiter with period below 3 days) are general features of secular chaos, or merely a consequence of our choice of initial conditions. We aim at answering these questions in the next section.

\section{Population Synthesis: exploring a wide range of initial conditions}
\label{sec:sims}

\subsection{Initial conditions}

In the previous section we varied only one control parameter, the semi-major axis of the outermost planet, to illustrate some of the key features of secular chaos, with and without dynamical tides. Here we focus only on secular chaos with dynamical tides, but consider a wide range of initial conditions and parameters to explore the outcome produced by a large number of simulations.

\begin{figure}
    \begin{center}
    \includegraphics[scale=0.85]{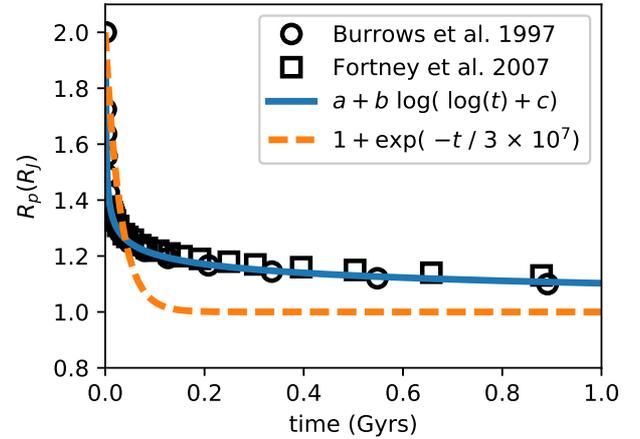}
    \caption{Time evolution of the radius of a giant planet at 1 au. Data taken from \citet{burrows97} and \citet{fortney07}. The best fit is the blue line with $a=1.656$, $b=-0.277$ and $c=-13.353$. It is valid for $t>10^6$ yr. For $t<10^6$ yr, we keep the radius of the planet fixed at $1.87\Rj$. The dashed orange line is the model used by \citet{wu07} and \citet{petrovich15a}.} 
    \label{fig:rptime}
    \end{center}
\end{figure}

In Table \ref{tab:ICall} we list the input parameters that we vary in our simulations. We used the fit from \citet{dk91} to compute the stellar radius from the stellar mass: $R_* = 1.06 (M_*/\Msun)^{0.945}\Rsun$. As we have seen in Section \ref{sec:ill}, the planet radius is a key parameter in determining the final orbital period of hot Jupiters. Gaseous planets are expected to cool down and shrink in time. To take this into account, we examine the evolutionary models of giant planets orbiting a Sun-like star at 1 au from \citet{burrows97} and \citet{fortney07}. We provide an analytic fit to these model results and use it in our simulations. This is shown in Figure \ref{fig:rptime}, along with a recipe that was used by \citet{wu07} and \citet{petrovich15a}. We set the initial radius of our inner planet to $1.87\Rj$, a value at which it stays for the first million years, until it decrease according to the fitting formula of Figure \ref{fig:rptime}. There is no need to consider the radius evolution of the outer two planets.

\begin{figure}
    \begin{center}
    \includegraphics[scale=0.5]{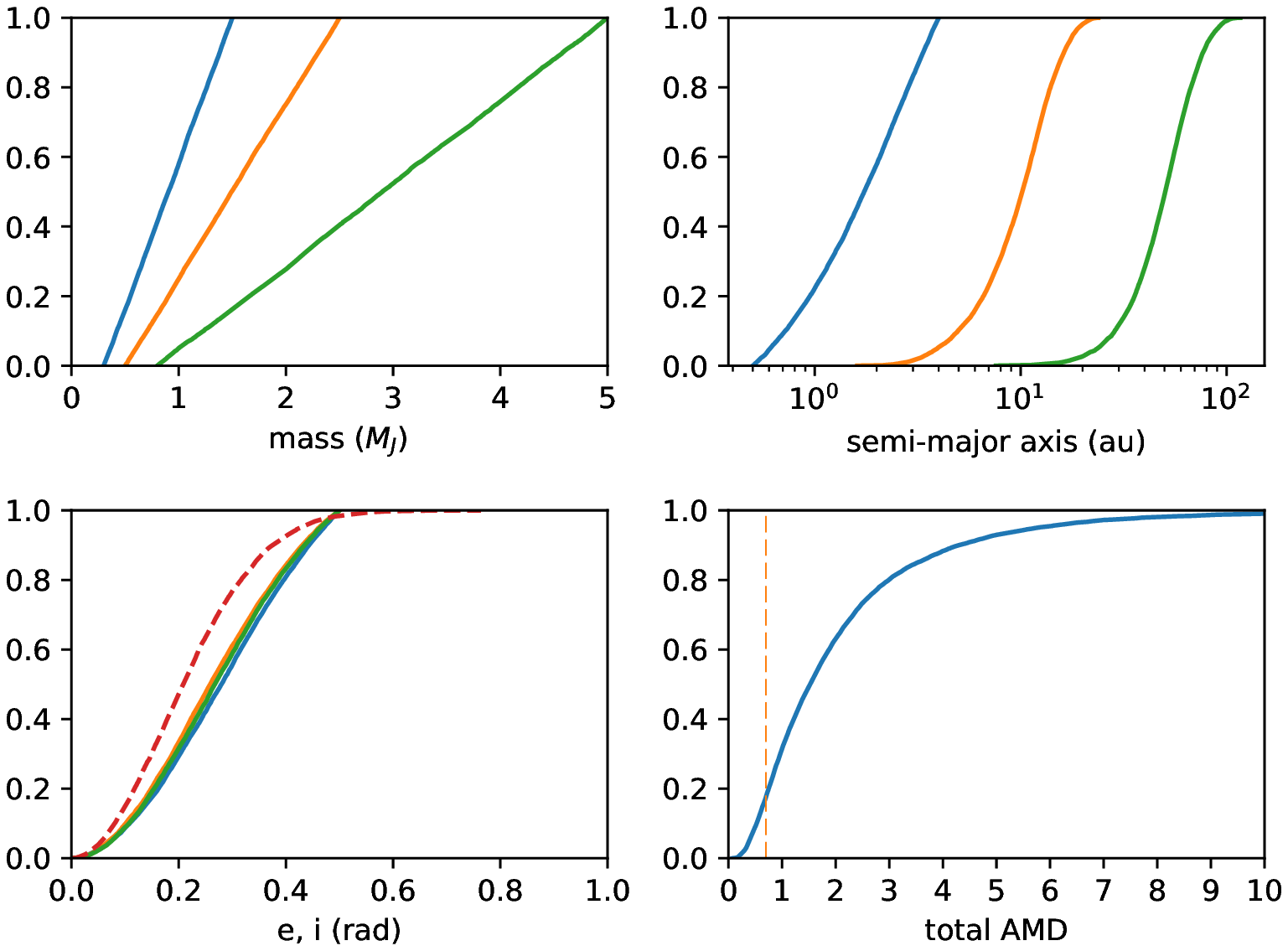}
    \caption{Planetary masses and initial orbits of the three planets for the population synthesis study of Section \ref{sec:sims}, represented as cumulative distribution functions. The blue, orange and green curves represent the innermost, middle and outermost planet, respectively. In the eccentricity and inclination plot (bottom left), the red dashed line is the inclination distribution, which is the same for all three planets. The bottom right panel is the total AMD distribution, in units of $\Lambda_1$ (see equation \ref{eq:lambda}). The vertical dashed line gives an estimate of the minimum amount of AMD required to form a hot Jupiter with no inclination (see equation \ref{eq:amdcrit}).} 
    \label{fig:ICall}
    \end{center}
\end{figure}

For initial conditions we require that the planets be well spaced. To this end, if the initial conditions are such that the pericenter of a planet is initially smaller than 1.5 times the apocenter of the planet immediately interior to it, we redraw the eccentricities and semi-major axis of all the planets. This means that the actual initial eccentricity distribution slightly deviates from the Rayleigh distribution. Figure \ref{fig:ICall} shows the distributions of mass, semi-major axis, eccentricity, inclination, and total AMD that we use in our simulations. The AMD distribution is such that most runs should in principle have enough AMD to produce a hot Jupiter, should chaotic diffusion take place and drive the eccentricity near unity. All other parameters are set to their fiducial values, and the integration time is 2 Gyr, as discussed in Section \ref{sec:num}.

\setlength{\tabcolsep}{4pt}
\begin{table}
  \caption{Initial distribution for the runs of Section \ref{sec:sims}.}
  \label{tab:ICall}
  \begin{tabular}{ll}
    \hline
    Parameter & Initial distribution  \\
    \hline
    $m_1$ & $[0.3-1.5]\Mj$  \\
    $m_2$ & $[0.5-2.5]\Mj$  \\
    $m_3$ & $[0.8-5.0]\Mj$  \\
    $a_1$ & $[0.5-2]$ au  \\
    $a_2$ & $[4-7]a_1$  \\
    $a_3$ & $[3-6]a_2$  \\
    $e$ & Rayleigh distrib. with $\sigma=0.27$, cutoff at $e=0.6$  \\
    $i$ & Rayleigh distrib. with $\sigma=7^{\circ}$, cutoff at $i=45^{\circ}$\\
    $\varpi$ & Uniform in $[0-2\pi]$ rad \\
    $\Omega$ & Uniform in $[0-2\pi]$ rad \\
    $M_*$ & Uniform in $[0.8-1.4]\Msun$ \\
    \hline
  \end{tabular}
\end{table}

\subsection{Results}
We carried out 7000 simulations as described above. Here we present the main results.

\subsubsection{Warm Jupiters properties}

\begin{figure}
    \begin{center}
    \includegraphics[scale=0.85]{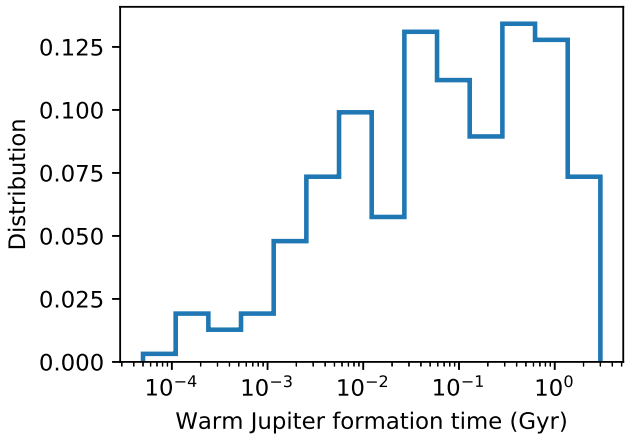}
    \caption{Distribution of the warm Jupiters formation time, defined as the time when the proto-hot Jupiter exits the dynamical tides phase.} 
    \label{fig:wj_time}
    \end{center}
\end{figure}

The hot Jupiter formation process is comprised of several stages. First the inner planet is ``chaotically'' pushed to high eccentricity due to planet-planet interactions; then it undergoes a phase of dynamical tides that quickly shrinks its semi-major axis and moderately circularizes the eccentricity (from $e\sim 0.99$ to 0.9) --- the result is a transient eccentric warm Jupiter; this is followed by gradual orbital decay and circularization due to weak friction tides, leading to the formation of a HJ. 
An example is shown in Figure \ref{fig:DTnoDT}. The transient warm Jupiters have semi-major axes that are a fraction of an au. 
Assuming angular momentum conservation, one can determine the final period of the hot Jupiter simply from the orbital parameters of the warm Jupiter formed  when chaotic tides are over. The warm Jupiter formation time is independent of the parameters that enter in the parametrisation of weak friction theory, and is therefore a more relevant time marker. Figure \ref{fig:wj_time} shows the distribution of the warm Jupiter formation time. The distribution rises steadily as a function of $\log$(time) for the first 100 million years, and then plateaus until 2 Gyrs  (which is our maximum integration time). By integrating longer we would have been able to create 10\% more warm Jupiters (and hence eventually more hot Jupiters.)

\begin{figure}
    \begin{center}
    \includegraphics[scale=0.85]{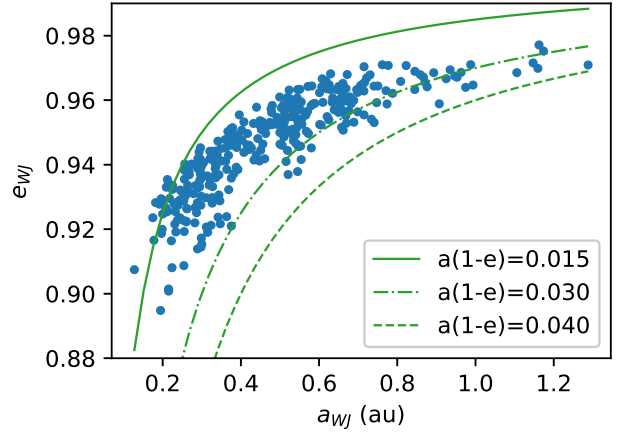}
    \caption{Semi-major axis and eccentricity of the transient warm Jupiters formed in the secular chaos scenario with dynamical tides. The green lines are lines of constant $a(1-e)$.} 
    \label{fig:wj_ae}
    \end{center}
\end{figure}

In Figure \ref{fig:wj_ae} we show the semi-major axis and eccentricity of the warm Jupiters. The initial semi-major axis distribution is uniform between 0.5 and 2 au, and the warm Jupiters are found to be closer to their host stars due to the shrinkage of semi-major axis caused by dynamical tides. The eccentricities are very high, meaning that weak friction tides can subsequently circularize the orbits and form hot Jupiters. 

\subsubsection{Hot Jupiters properties}

Table \ref{tab:results} summarizes the formation and disruption fractions of HJs in our population synthesis calculations of three-planet systems. With our set of initial conditions, $4.5$\% of all inner planets ended up as a hot Jupiter without being disrupted (recall that we consider a planet to be disrupted when its pericenter comes within $2\rt$). When applying a cut-off at $2.7\rt$, the fraction of hot Jupiters drops to $2.1$\%. A small number of runs were stopped because disruption of the inner planet occurred, or because two planets came close enough to each others that we considered the secular approximation not to be valid any more (see Section \ref{sec:stab}). 

\begin{figure*}
    \begin{center}
    \includegraphics[scale=0.85]{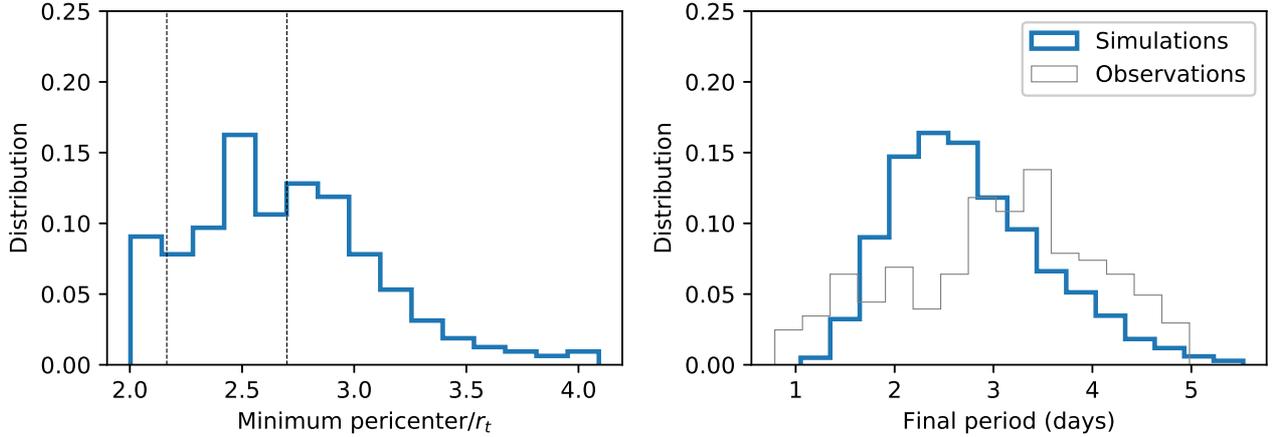}
    \caption{Results of our ensemble of simulations described in Section \ref{sec:sims}. \textit{Left:} distribution of minimum pericenter distances (in units of tidal radius, see eq. \ref{eq:rt}). The two vertical lines show the disruption radii at 2.16 and 2.7 $\rt$, respectively. Planets to the left of these lines could be disrupted. \textit{Right:} distribution of final orbital periods. The blue histogram is the result of our simulations while the grey histogram is the observed distribution for all planets with $M\sin i>0.3~\Mj$ and period below 5 days (data taken from exoplanets.org). 
    } 
    \label{fig:hj_distrib}
    \end{center}
\end{figure*}

In Figure \ref{fig:hj_distrib} we examine more closely the properties of the hot Jupiters formed in our population study. First we confirm that the choice of tidal disruption radius has a major impact on the survival rate. Choosing a threshold value of $2.7\rt$ for disruptions would mean that roughly 50\% of the hot Jupiter formed in our simulations were disrupted in the process of migrating, while a value of $2.16\rt$ reduces this number to 5\%. Comparing the HJs produced in our simulations with the observed systems, we find that the final period distribution of our HJs peaks between 2 and 3 days, while the observations show a peak a 3 to 4 days. As shown in Section \ref{sec:ill}, the planet radius is an important factor in setting the final period \citep[see also][]{wu07}. 

Because secular chaos operates on a timescale similar or longer than the planet radius shrinking timescale, our proto-hot Jupiters enter the phase of dynamical tides with a radius that is already too small to allow for large-period planets to be formed. To check this we have rerun a smaller series of simulations using the same radius evolution formula as \citet{wu07}. 
We find an overall similar final period distribution (not shown here). Although the long-term radius shrinkage is rather different between the two fitting formulas, the initial rapid decrease from a large radius to about $1.2\Rj$ occurs within the first 100 million years in both cases. Since this corresponds to the time where the bulk of our hot Jupiters are formed (see Figure \ref{fig:wj_time}), it is not surprising that the differences in the outcome are only minor.

\begin{figure*}
    \begin{center}
    \includegraphics[scale=0.85]{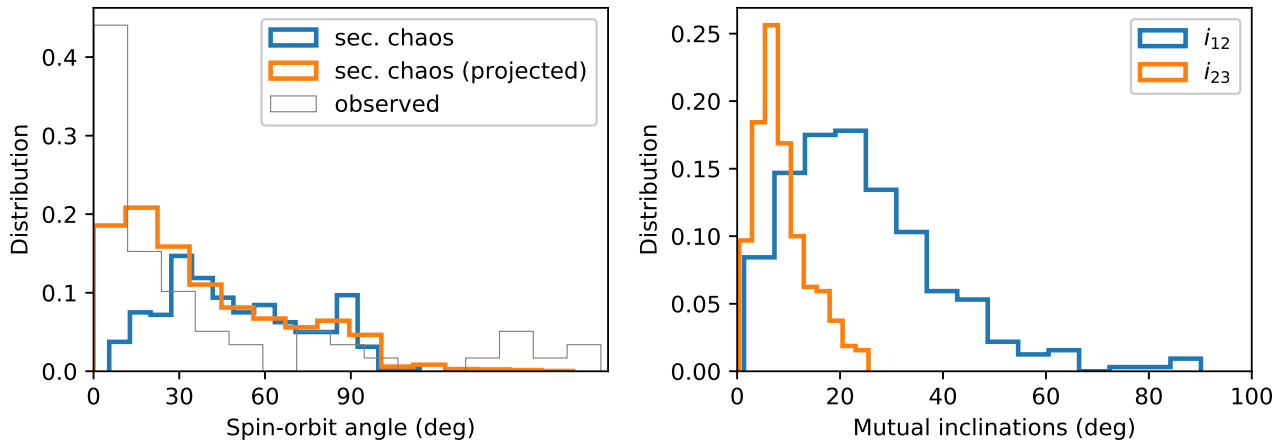}
    \caption{Various angle distributions for our ensemble of simulations described in Section \ref{sec:sims}. \textit{Left:} The blue line is the distribution of real spin-orbit angles from the simulations, while the orange line is the projected distribution. The grey line is the observed projected distribution. \textit{Right:} Distribution of mutual inclinations between planets 1 and 2 ($i_{12}$) and between 2 and 3 ($i_{23}$).
    } 
    \label{fig:hj_inc}
    \end{center}
\end{figure*}

In Figure \ref{fig:hj_inc}, we show the spin-orbit angle distribution for all the hot Jupiters formed in our simulations. The observed distribution of spin-orbit angles is in fact for the sky-projected angles. To convert from our true distribution to a projected distribution, we used the procedure of \citet{fw09}, assuming that the observer sees the transit exactly edge on. We find a lack of very aligned planets. This discrepancy could be resolved if some of the aligned hot Jupiters have formed by disc migration, or by coplanar high-eccentricity migration \citep{petrovich15b}. In addition, we find that secular chaos produces a lack of retrograde planets compared to observations \citep[see also][]{lw14}. In Figure \ref{fig:hj_inc} we also show the mutual inclinations between adjacent pairs of planets. The distribution of inclination between the hot Jupiter and planet 2 ($i_{12}$) is very broad, while the mutual inclination of the outer two planets is small, as most of the AMD of the outer planets has diffused and gone into the innermost planet.

\begin{figure}
    \begin{center}
    \includegraphics[scale=0.85]{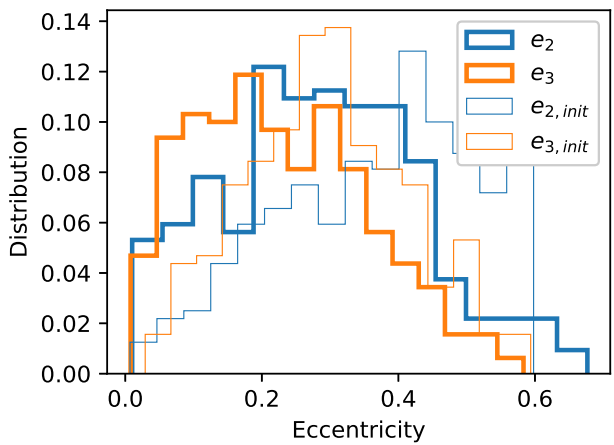}
    \caption{Initial (thin) and final (thick) distributions of eccentricities for planet 2 (blue) and 3 (orange) of all systems that form a hot Jupiter. 
    } 
    \label{fig:hj_e2e3}
    \end{center}
\end{figure}

Since it might prove useful when comparing with actual systems containing a hot Jupiter and a planetary companion at a few au's, we show in Figure \ref{fig:hj_e2e3} the eccentricity distribution for the outer two planets in all systems that form a hot Jupiter. For comparison, the initial eccentricity distribution of these systems is also shown. Note that the distribution of initial eccentricities does not match the distribution of our entire population shown in Figure \ref{fig:ICall}. This is because only a sub-sample of the initial systems lead to the formation of hot Jupiters. Figure \ref{fig:hj_e2e3} shows that systems where the outer two planets have large eccentricities favour the formation of a hot Jupiters. This is expected as these systems contain more AMD. According to the theory of secular chaos, these AMD diffuse from the outer planets to the inner one in order to form a hot Jupiter. We therefore expect planet 2 and 3 to have lost a significant amount of AMD (and therefore eccentricity) in the process of forming a hot Jupiter. Figure \ref{fig:hj_e2e3} shows that this is indeed the case.

\begin{figure}
    \begin{center}
    \includegraphics[scale=0.85]{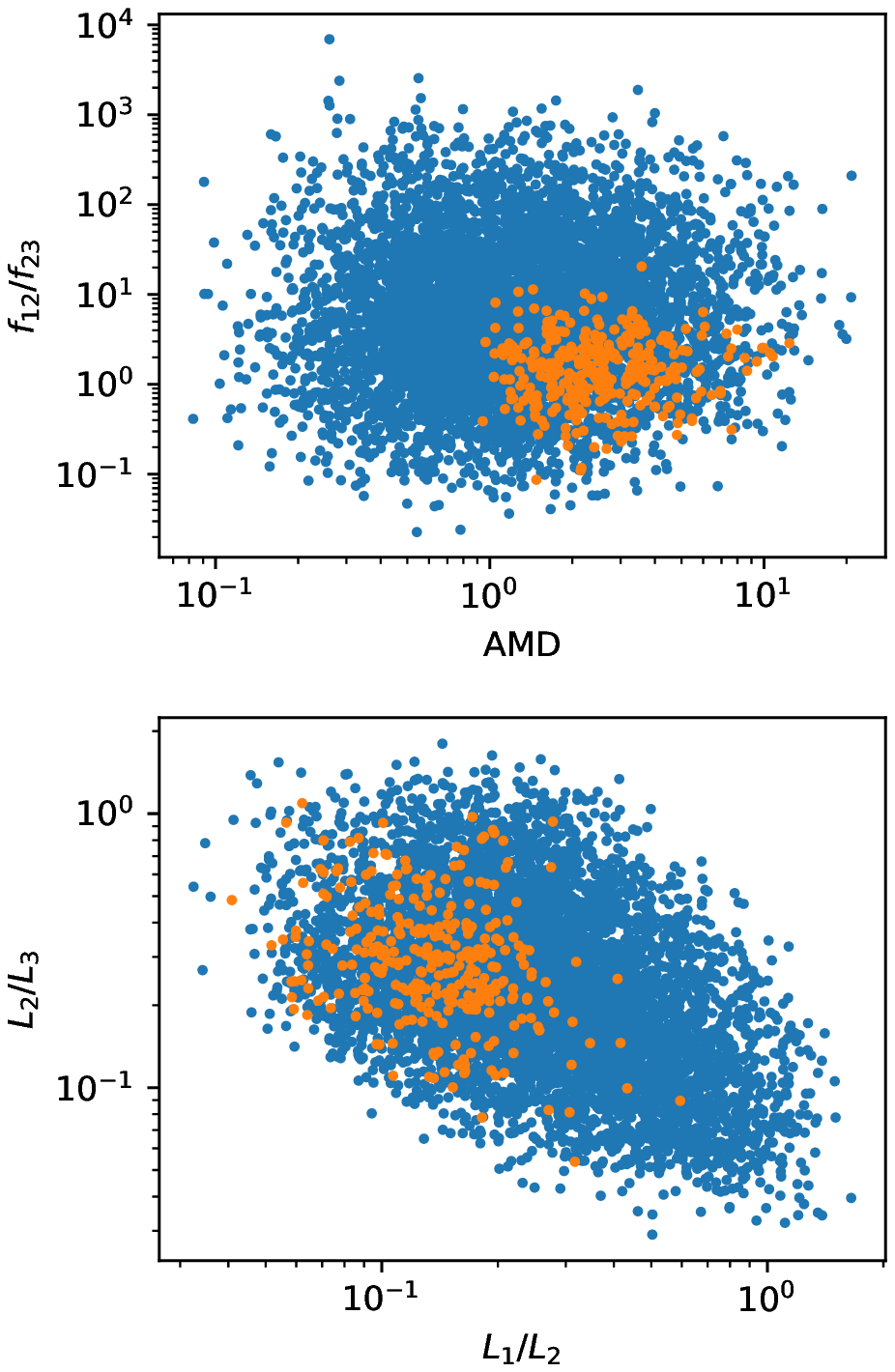}
    \caption{Various cuts in parameter space showing what initial configurations are more likely to form a hot Jupiter. Orange dots are simulations that formed a hot Jupiter while blue dots are simulations that did not form a hot Jupiter. The top panel shows that the rate of precession induced by planet 3 on 2 ($f_{23}$) needs to be close to that induced by 2 on 1 ($f_{12}$), but also that the AMD of the system needs to be equal to or larger than $\Lambda_1$. The bottom panel shows the ratio of angular momentum of planets 1 to 2 and 2 to 3. Our initial parameter space is such that in general, planet 3 carries most of the angular momentum. Although there is no strong preference in $L_2/L_3$, it is clear that secular chaos is most efficient when $L_1\ll L_2$.} 
    \label{fig:hj_param}
    \end{center}
\end{figure}

Finally, it is worth asking whether, despite the chaotic evolution of the systems, one can identify a preferred region of parameter space from which the proto-hot Jupiters originate. The parameter space of a 3-planet system is huge, and one needs to find the relevant quantities that capture the dynamics of the system. As discussed in Section \ref{sec:secres}, large AMD and closeness to a secular resonance are important. Let's simplify even further the toy model we have laid down in Section \ref{sec:secres} and define
\begin{align*}
f_{12} &= \frac{m_2}{\Ms}\left(\frac{a_1}{a_2}\right)^{3} n_1,\\ 
f_{23} &= \frac{m_3}{\Ms}\left(\frac{a_2}{a_3}\right)^{3} n_2.
\end{align*}
When $L_1\lesssim L_2\lesssim L_3$, these provide estimates for the the precession frequency induced by planet 2 on 1, and planet 3 on 2, respectively. We expect that systems with $f_{12}\sim f_{23}$ would be close to a secular resonance and therefore more prone to secular chaos. On the top panel of Figure \ref{fig:hj_param} we show that indeed most hot Jupiters originate from the region of parameter space where $f_{12}\sim f_{23}$ and where the AMD is larger than $\Lambda_1$. This agrees with our predictions in Section \ref{sec:sechaos} regarding the onset of secular chaos. Since our toy model was derived with the assumption that the inner planet could be treated as a test particle, we also show on the bottom panel of Figure \ref{fig:hj_param} the ratio of initial angular momentum of planet 1 to 2 versus 2 to 3. Our initial conditions are such that we only explore a certain region of this parameter space. However it is clear that while secular chaos seems to be able to operate for a large range of values of $L_2/L_3$, it only works if $L_1\ll L_2$. This explains why our toy model discussed in Section \ref{sec:secres} gave accurate predictions regarding the onset of secular chaos and the location of secular resonances.

\begin{table}
  \centering
  \begin{tabularx}{0.85\columnwidth}{@{\extracolsep{1cm}}lc}
  	\hline
  	Outcome & Fraction \\
  	\hline
    hot Jupiters with $\min(p)>2.\rt$ &  4.5\% \\
    hot Jupiters with $\min(p)>2.7\rt$ &  2.1\% \\
    disrupted planets $(\min(p)<2\rt)$ & 1.5\% \\
    near-crossing orbits & 5.2\% \\ 
	\hline
  \end{tabularx}
  \caption{Results out of 7000 runs of 3-planet systems described in Section \ref{sec:sims}. Here $\min(p)$ is the minimum pericenter distance reached by the innermost planet. Near-crossing orbits mean that two planets came too close to each others and the integration was stopped (see eq. \ref{eq:stabP15}).}
  \label{tab:results}
\end{table}

\section{Discussion}
\label{sec:discussion}

Here we discuss a few points that are worth considering when interpreting our results and that of others in the literature. 

In this paper, using the secular ring code, we were able to integrate a large ensemble of 3-planet systems to systematically study the outcome of secular chaos. Ideally, in order to obtain the steady-state rate of hot Jupiter formation, one should carry on numerical integrations for a time drawn randomly between 0 and, say, 10 Gyrs. In practice, we found that the computational cost of accurately integrating a 3-planet system for 10 Gyrs (accurate in the sense of AMD conservation) was prohibitive, and we limited ourselves to 2 Gyr-long integrations. Given the large parameter space, our goal was to explore the outcome of secular chaos and dynamical tides rather than to construct a population synthesis. The distribution of the HJ formation time shown in Figure \ref{fig:wj_time} suggests that the bulk of the hot Jupiters formed early on, and integrating over 10 Gyrs would only modify our results only slightly.

We have found that the formation of hot Jupiters through secular chaos shares some common features with the Lidov-Kozai mechanism \citep{wu18,vla18}, but also some major differences. For example, in both scenarios, dynamical tides can save the planet from tidal disruption under some conditions. The main difference is that secular chaos is often a ``one shot'' process. That is, one single phase of high eccentricity is usually what is needed to form a hot Jupiter (or disrupt the planet). In contrast, the Lidov-Kozai simulations of \citet{vla18} showed that the planet often goes through multiple episodes of high eccentricity, during which dynamical tides gradually reduce the semi-major axis in a series of steps. It was found that this pathway yields longer period planets that make up the tail of the predicted HJ period distribution. Another consequence of these multiple episodes of chaotic tides in Lidov-Kozai oscillations is that the planet has more than one opportunity to significantly alter the spin axis of the star. It was indeed shown by \citet{storch14} and \citet{anderson16} that the evolution of the stellar spin can be chaotic as the planet undergoes Lidov-Kozai migration, and large spin variations can be produced, thus producing a wide range of spin-orbit misalignments. It was our hope when starting this work that a similar strongly chaotic spin evolution would happen in secular chaos, helping us overcome the issue observed by \citet{lw14}  (who did not include spin-orbit coupling) that spin-orbit angles larger than 90 degrees are hard to generate. This is unfortunately not the case due to the ``one shot'' nature of secular chaos.

Despite the large number of simulations in Section \ref{sec:sims}, we should consider this population synthesis model with caution because of the uncertainty in the initial conditions. At this point we have no way to justify why a system of three giant planets should follow the initial conditions we have used. We could speculate	on the origin of AMD in planetary systems. One scenario that could lead to our initial conditions is to assume that the system went through a phase of violent dynamical relaxation in its very early age, just after the proto-planetary disc had dispersed, a phenomenon usually referred to as planet-planet scattering \citep[see, e.g.,][]{rf96,pt01,chatterjee08,jt08}. Multiple close approaches between giant planets lead to ejections or collisions, and simulations show that the systems are often left with two or three planets on eccentric and inclined orbits. One may also go even one step further in the past history of the system, and consider planet-disc interactions. How the AMD of a coupled disc-planet(s) system evolves is still unknown. However, \citet{lo01} (for inclinations) and \citet{to16} (for eccentricities) have shown that Lindblad resonances can excite the AMD on secular timescales when considering giant, gap-opening, planets. Simulations by \citet{ragusa18} have shown that the eccentricity of giant planets can indeed grow on such timescales, although more work is needed to explore the mass regime of jovian planets. 
 
\section{Conclusion}
\label{sec:conclu}

In this paper we have presented a comprehensive study of hot Jupiter
formation in the ``secular chaos'' high-eccentricity migration
scenario first proposed by \citet{wl11}. In this scenario,
secular interactions among three giant planets (``cold Jupiters'')
chaotically excite the inner planet's eccentricity, pushing
it to highly eccentric orbits, sufficient to form hot Jupiters via
subsequent tidal circularization and orbital decay. Using a Gaussian
ring secular code and incorporating dynamical tides from
\citet{vla18}, we carry out calculations for a large ensemble of
systems to explore the conditions for secular chaos to excite extreme
eccentricities and the outcomes of this hot Jupiter formation channel.

Our calculations show that the parameter space for hot Jupiter
formation can be understood by requiring (i) the AMD be sufficiently
large, and (ii) there exist a secular resonance for a range of inner
planet eccentricities (Section \ref{sec:sechaos}; see Figs. \ref{fig:a3freq}, \ref{fig:a3pmin} and \ref{fig:hj_param}). The secular
equations based on Gaussian rings (and including various short-range
forces) capture the essential feature of the secular chaos of 3-planet
evolution; moreover, they allow for fast computations of the long-term
(2~Gyrs) orbital evolution for a large number of systems that are
otherwise impractical with $N$-body code. The results from our
simulations show that without the inclusion dynamical tides, most
planets that experience large eccentricity excitation would be tidally
disrupted before they can form a hot Jupiter (see Fig. \ref{fig:a3pmin}a), especially
if we adopt the more conservative disruption criterion (i.e. tidal
disruption occurs when the pericenter distance is less than $r_{\rm dis}
=2.7\rt$, with $\rt=R_p(M_\star/M_p)^{1/3}$)
suggested by hydrodynamical simulations.
With dynamical tides, however, planets can be ``saved'' from tidal
disruption (see Figs. \ref{fig:DTnoDT} and \ref{fig:a3pmin}b) and become eccentric warm Jupiters
(see Fig. \ref{fig:wj_ae}), which then circularize by equilibrium tides to form hot
Jupiters.  The final periods of these hot Jupiters lie predominantly
around 2-3 days (see Fig. \ref{fig:hj_distrib}), somewhat shorter than the observed
periods (3-4 days).
We include spin-orbit coupling (between the stellar spin axis and the
planet's orbital axis) and stellar spin down due to magnetic braking
in our secular evolution. Although these physical effects strongly 
influence the stellar spin evolution, we find that the final spin-orbit 
misalignments are predominantly prograde ($\theta_{\rm sl}<90^\circ$), and
almost no retrograde systems are produced (see Fig. \ref{fig:hj_inc}).
Taken together, the high tidal disruption rates, small orbital periods
and lack of retrograde spin-orbit misalignments suggest that
high-eccentricity migration via secular chaos can only account for a
fraction of the observed hot Jupiter population.

There remain a number of uncertainties in our modelling of hot Jupiter
formation via secular chaos. Besides the uncertainty in the initial
conditions of the multi-planet systems (see Section \ref{sec:discussion}), the most
important are tidal disruption and tidal dissipation. Our work has
highlighted the critical role of tidal disruption radius: The current
estimate ranges from $2.16r_{\rm tide}$ to $2.7r_{\rm tide}$, and the
precise value makes a large difference in the hot Jupiter formation efficiency.
Future work should also address the effects that energy deposition and
tidal heating have on the upper layers and internal structure of the
planet, especially since the planet radius appears to play an
important role in the finale period distribution.

\section*{Acknowledgements}
We thank Will M. Farr for making his code publicly available, and
Kassandra Anderson and Bonan Pu for fruitful discussions.  This work
is supported in part by the NSF grant AST1715246 and NASA grant
NNX14AP31G. MV is supported by a NASA Earth and Space Sciences
Fellowship in Astrophysics.

\bibliographystyle{mnras}
\bibliography{biblio}

\appendix

\section{Stability criterion}
\label{app:stab}

\begin{figure*}
    \begin{center}
    \includegraphics[scale=0.6]{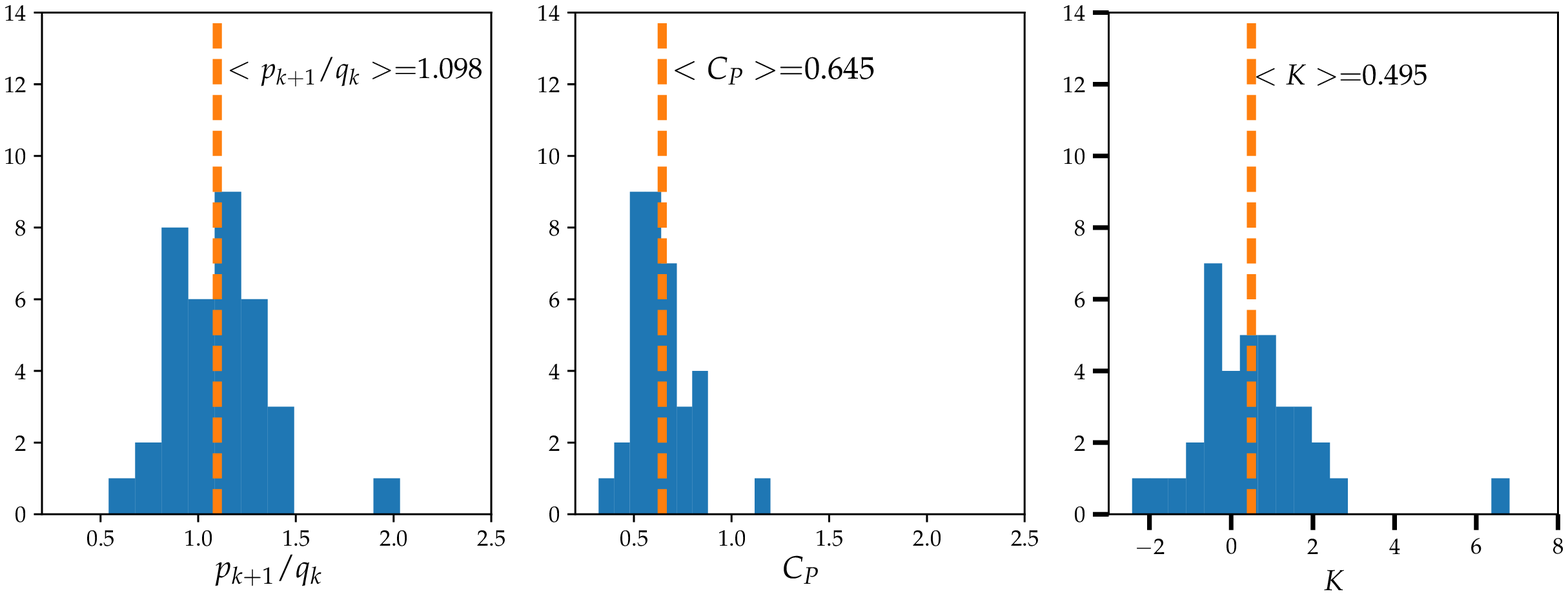}
    \caption{Left panel: Distribution of the ratio of apocenter to pericenter of two adjacent planets at the time when systems undergo a semi-major axis change of more than 5\%. Middle panel: same, but showing the distribution of $C_P$ instead (eq. \ref{eq:stabP15}). Right panel: same, but showing the distribution of the mutual separation in Hill radii instead (eq. \ref{eq:stabHill}). In each panel, the vertical orange dashed line indicates the mean of the distribution.} 
    \label{fig:SCstab}
    \end{center}
\end{figure*}

In this appendix we discuss how we derived a criterion from $N$-body simulations that can help identify systems that would potentially go unstable in secular codes.

We ran a sample of $N$-body simulations using the {\sc mercury6} Bulirsh-Stoer integrator with an accuracy tolerance of $10^{-12}$ and initial time-step of 2 days. No tidal dissipation was included but an extra force accounting for relativistic precession was added. Our simulations contained three planets orbiting a solar-mass star. The initial conditions for the three planets are given in Table \ref{tab:ICmercury}. The outermost planet's mass and semi-major axis were drawn uniformly between 1.5 and 2.8 $\Mj$ and 14 and 24 au, respectively. Once this was done, we randomly picked the criterion $C_{\rm P}$ from Equation (\ref{eq:stabP15}) between 0.5 and 1.5, and deduced the eccentricity of the outermost planet. The total integration time was up to $10^9$ yrs. Integrations were stopped when: i) a planet hit the star or ii) the semi-major axis of a planet changed by more than 5\%. This last condition is meant to represent a breakdown of the secular theory, which in principle conserves energy and therefore semi-major axis. We then analysed the sample of unstable runs, which consists of 50 simulations. We kept track of the last output before the run was labeled as unstable. 

From this $N$-body integration data, we attempted to derive a relatively simple criterion for unstable systems, that can be implemented in a secular code. Perhaps the simplest approach would consist of measuring the ratio of pericenter $p_{k+1}$ to apocenter $q_k$ of two adjacent planets labelled by $k$. We can define a stability criterion by the typical value of $p_{k+1}/q_k$ at which two planets become unstable.
Another approach could be to use a modified mutual Hill radius for two planets, of the form
\begin{equation}
r_H=\left(\frac{m_k+m_{k+1}}{3M_*}\right)^{1/3}\frac{q_k+p_{k+1}}{2},
\end{equation}
and the dimensionless separation
\begin{equation}
\label{eq:stabHill}
K=\frac{a_{k+1}-a_k}{r_H},
\end{equation}
and again ask for which value of $K$ does the system become unstable.

The last criterion we check is the one of \citet{petrovich15c} (see Eq. \ref{eq:stabP15}).
It is an extension of the ratio of pericenter to apocenter criterion, that includes a dependence on the mass and semi-major axis of the adjacent planets as well. In Petrovich's simulations, systems with initial $C_{\rm P}<1$ are likely to be unstable.

The results are shown in Figure \ref{fig:SCstab}. For each of the three metrics that we considered, we find a similarly shaped distribution which centres around a single mean. For the ratio of pericenter to apocenter, the mean ratio is about 1.1, for the Petrovich criterion it is 0.65, and for the mutual separation it is 0.5 Hill radii. Note the presence of one outlier in all distributions, which corresponds to the same system. 

Based on these simulations we decided to consider our secular code to become invalid (and therefore stop the integration) when $C_{\rm P}<0.75$.

Note that Petrovich's criterion is based on the initial conditions of the system, while here we ask a slightly different question: in a 3-planet system, how close can two planets get before the secular approximation breaks down (i.e., before semi-major axis variations become significant). We stress here that the variables that go into our criterion $C_{\rm P}<0.75$ are dynamical variables that evolve in time, not initial conditions. The criterion is therefore evaluated at each timestep of the secular code. Our $N$-body simulations was simply designed to get a rough idea of when the breakdown happens, and further work would be needed to rigorously assess the stability of such systems.

\setlength{\tabcolsep}{4pt}
\begin{table}
  \caption{Initial conditions for three planets}
  \label{tab:ICmercury}
  \begin{tabular}{lcccccc}
    \hline\hline
    Planet &  mass ($\Mj$) & $a$ (AU) & $e$ & $i$ (deg) & $\omega$ (rad) & $\Omega$ (rad)\\
    \hline
    1 & 0.8 & 1.0 & 0.066 & 4.5 & $\pi$ & 0   \\ 
    2 & 1.3 & 6.0 & 0.18 & 19.9 & $0.38$ & $\pi$   \\ 
    3 & [1.5-2.8] & [14-24]& see text & 7.9 & $\pi$ & 0   \\ 
    \hline
  \end{tabular}
\end{table}

\section{Additional information on tides}
\label{app:tides}

\subsection{Weak friction theory}
\label{app:wft}

We give the expression for the functions $f_1$ to $f_5$ that appear in the weak friction theory:
\begin{align}
f_1(e) &= 1 + \frac{31}{2}e^2 + \frac{255}{8}e^4 + \frac{185}{16}e^6 + \frac{25}{64}e^8, \\
f_2(e) &= 1 + \frac{15}{2}e^2 + \frac{45}{8}e^4 + \frac{5}{16}e^6, \\
f_3(e) &= 1 + \frac{15}{4}e^2 + \frac{15}{8}e^4 + \frac{5}{64}e^6, \\
f_4(e) &= 1 + \frac{3}{2}e^2 + \frac{1}{8}e^4, \\
f_5(e) &= 1 + 3e^2 + \frac{3}{8}e^4.
\end{align}

\subsection{Dynamical tides}
\label{app:dt}

The energy transfer $\Delta E$ between the mode and a parabolic orbit was first studied by \citet{pt77}, and then generalized to eccentric orbits and to include the effect of rotation \citep{lai97,fl12}. For a planet with pericenter distance $r_{\rm p}$ arund a star $M_\star$.  the energy transfer to the planetary modes (keeping only the modes with degree l=2) reads
\begin{equation}
\Delta E = \frac{G\Ms^2}{r_{\rm p}^6} \Rp^5 T(\eta,\sigma/\Omega_{\rm peri},e).
\end{equation}
Here $\eta=r_{\rm p}/r_{\rm t}$, the ratio of pericenter to tidal radius (see eq. \ref{eq:rt}), and $\Omega_{\rm peri}=(GM_\star/r_p^3)^{1/2}$. The dimensionless function $T$ is given by
\begin{equation}
T=2\pi^2 \frac{\sigma}{\epsilon}Q^2K_{lm}^2.
\end{equation}
We only consider the dominant prograde f-mode with $l=m=2$, and we assume that the planet is composed of a neutrally stratified fluid described by a $\gamma=2$ polytrope. In the above formula, $\sigma$ is the mode frequency in the inertial frame (i.e., given a mode with mode number $m$ and frequency in the rotating frame $\omega$, we have $\sigma=\omega+m\Omega_p$), and $\epsilon$ is a frequency that only differs from $\omega$ by a correction due to rotation (in the slowly-rotating approximation). In the slowly-rotating approximation, we have $\epsilon=1.22 (G\Mp/\Rp^3)^{1/2}$, $\omega=\epsilon-\Omega_{\rm p}$, and $\sigma=\epsilon+\Omega_{\rm p}$. In addition, $Q$ is an overlap integral, whose numerical value is 0.56 for our specific model, and \citet{lai97} gave an approximation for $K_{22}$ accurate within 2\% for $(1-e)\ll 1$ and $z\equiv\sqrt{2}\sigma/\Omega_{\rm peri}$ larger than a few:
\begin{equation}
K_{22}=\frac{2z^{3/2}\exp(-2z/3)}{\sqrt{15}}\left(1-\frac{\sqrt{\pi}}{4\sqrt{z}} \right).
\end{equation}
More details can be found in \citet{vla18}.

\end{document}